\documentclass[twocolumn]{IEEEtran}

\usepackage{amsmath, amssymb, latexsym, amsthm, graphicx, epsfig, subfigure, cite, pifont, xcolor, scalerel, enumerate, multicol, lipsum, cuted, stfloats, booktabs, setspace, amsbsy}
\usepackage[ruled, linesnumbered]{algorithm2e}

\newcommand{\bmby}{\bar{\mathbf{y}}}
\newcommand{\mby}{\mathbf{y}}
\newcommand{\mbh}{\mathbf{h}}
\newcommand{\mbw}{\mathbf{w}}
\newcommand{\mbX}{\mathbf{X}}
\newcommand{\mbx}{\mathbf{x}}

\newcommand{\mbs}{\mathbf{s}}

\newcommand{\mbF}{\mathbf{F}}
\newcommand{\mbK}{\mathbf{K}}
\newcommand{\mbI}{\mathbf{I}}
\newcommand{\mbq}{\mathbf{q}}
\newcommand{\mbQ}{\mathbf{Q}}

\newcommand{\mbFA}{\mathbf{F}^\mathsf{A}}
\newcommand{\mbFP}{\mathbf{F}^\mathsf{P}}
\newcommand{\mbf}{\mathbf{f}}

\newcommand{\mbhA}{\mathbf{h}^\mathsf{A}}
\newcommand{\mbhP}{\mathbf{h}^\mathsf{P}}
\newcommand{\mbhAP}{\mathbf{h}^\mathsf{A,P}}
\newcommand{\hmbh}{\hat{\mathbf{h}}}
\newcommand{\hmbhAP}{\hat{\mathbf{h}}^\mathsf{A,P}}
\newcommand{\cS}{\mathcal{S}}
\newcommand{\ucS}{\underline{\mathcal{S}}}
\newcommand{\sfA}{\mathsf{A}}
\newcommand{\sfP}{\mathsf{P}}

\newcommand{\hAP}{h^\mathsf{A,P}}
\newcommand{\mbPsi}{\mathbf{\Psi}}

\newcommand{\mbpsi}{\boldsymbol{\psi}}

\newcommand{\mbxi}{\boldsymbol{\xi}}
\newcommand{\mblamb}{\boldsymbol{\lambda}}

\newcommand{\mbGam}{\boldsymbol{\Gamma}}
\newcommand{\mbPi}{\boldsymbol{\Pi}}

\DeclareMathOperator*{\argmin}{arg\,min}

\allowdisplaybreaks

\begin{document}

\title{Pilot Signal and Channel Estimator Co-Design for Hybrid-Field XL-MIMO}

\author{\IEEEauthorblockN{Yoonseong Kang, \IEEEmembership{Graduate Student Member, IEEE}, Hyowoon Seo, \IEEEmembership{Member, IEEE}, and Wan Choi, \IEEEmembership{Fellow, IEEE}}
\thanks{Yoonseong Kang is with the School of Electrical Engineering, Korea Advanced Institute of Science and Technology (KAIST), Daejeon 34141, Korea (e-mail: yoonseongkang@kaist.ac.kr).}
\thanks{Hyowoon Seo is with the Department of Electronics and Communications, Kwangwoon University, Seoul 01897, Korea (e-mail: hyowoonseo@kw.ac.kr).}
\thanks{Wan Choi is with the Department of Electrical and Computer Engineering and the Institute of New Media and Communications, Seoul National University (SNU), Seoul 08826, Korea (e-mail: wanchoi@snu.ac.kr) (Corresponding authors: Hyowoon Seo and Wan Choi.)}
}

\maketitle
\vspace{-0.3in}
\begin{abstract}
This paper addresses the intricate task of hybrid-field channel estimation in extremely large-scale MIMO (XL-MIMO) systems, critical for the progression of 6G communications. Within these systems, comprising a line-of-sight (LoS) channel component alongside far-field and near-field scattering channel components, our objective is to tackle the channel estimation challenge. We encounter two central hurdles for ensuring dependable sparse channel recovery: the design of pilot signals and channel estimators tailored for hybrid-field communications. To overcome the first challenge, we propose a method to derive optimal pilot signals, aimed at minimizing the mutual coherence of the sensing matrix within the context of compressive sensing (CS) problems. These optimal signals are derived using the alternating direction method of multipliers (ADMM), ensuring robust performance in sparse channel recovery. Additionally, leveraging the acquired optimal pilot signal, we introduce a two-stage channel estimation approach that sequentially estimates the LoS channel component and the hybrid-field scattering channel components. Simulation results attest to the superiority of our co-designed approach for pilot signal and channel estimation over conventional CS-based methods, providing more reliable sparse channel recovery in practical scenarios.
\end{abstract}

\begin{IEEEkeywords}
    Extremely large-scale MIMO, hybrid-field channel, pilot signal design, channel estimation, Bayes methods.
\end{IEEEkeywords}

\IEEEpeerreviewmaketitle

%%%%%%%%%%%%%%%%%%%%%%%%%%%%%%%%%%%%%%%%%%%%%%%%%%%%%%%%%%%%%%%%%%%
% % % % % % % % % % % % % % % % % % % % % % % % % % % % % % % % % %
%%%%%%%%%%%%%%%%%%%%%%%%%%%%%%%%%%%%%%%%%%%%%%%%%%%%%%%%%%%%%%%%%%%
\section{Introduction}
%%%%%%%%%%%%%%%%%%%%%%%%%%%%%%%%%%%%%%%%%%%%%%%%%%%%%%%%%%%%%%%%%%%
% % % % % % % % % % % % % % % % % % % % % % % % % % % % % % % % % %
%%%%%%%%%%%%%%%%%%%%%%%%%%%%%%%%%%%%%%%%%%%%%%%%%%%%%%%%%%%%%%%%%%%

In the transition from the fifth-generation (5G) to the upcoming sixth-generation (6G) era, the pursuit of significantly higher data rates, potentially exceeding 1 Tb/s, has become a central focus for both the industry and academia \cite{Andrews2014, Giordani2020, Kang2020, Lee2022, Lee2023}. One of the key technologies for achieving this ambitious goal is the implementation of an extremely large-scale MIMO (XL-MIMO) system, where a base station (BS) deploys extremely large antenna arrays to enhance spectral efficiency \cite{Carvalho2020}. Moreover, high-frequency band systems, such as millimeter-wave or sub-terahertz bands, offer the advantage of abundantly available bandwidth. Consequently, high-frequency XL-MIMO systems are envisioned as a fundamental means of achieving the enhanced spectral efficiency targeted by 6G. However, the integration of massive antenna arrays and high-frequency bands faces fundamental changes in the operational environments of the wireless communication system.

The majority of these changes are related to the properties of electromagnetic fields and are primarily driven by two factors: the increasing size of antenna apertures and the decreasing wavelength of signals. The electromagnetic field can be divided into two distinct regions widely known as far-field and near-field. The boundary separating these regions is defined by the Rayleigh distance, expressed as $Z = 2D^2/\lambda$, where $D$ represents the antenna diameter, and $\lambda$ is the wavelength \cite{Sherman1962}. For instance, in a traditional massive MIMO system with a 0.1-meter uniform linear array (ULA) operating at a carrier frequency of 30 GHz, the Rayleigh distance is only 2 meters. In contrast, in an XL-MIMO system with a 0.5-meter ULA at 100 GHz, the radiating near-field distance can extend up to 167 meters. Consequently, high-frequency XL-MIMO systems have the potential to operate in both the far-field and near-field regions, a departure from the conventional systems that typically function exclusively in the far-field region.

Due to the increased Rayleigh distance in high-frequency XL-MIMO systems, the conventional plane wavefront assumption that holds in the far-field region is no longer valid. Therefore, there is an imperative need to shift from the conventional plane wavefront assumption to a more precise spherical wavefront assumption in high-frequency XL-MIMO systems \cite{Zhang2023}. The spherical wavefront assumption, encompassing both the angle of arrival and the distance between the transmitter and receiver in the array steering vector, facilitates the generation of focused beams in the near-field region, a phenomenon known as beam-focusing. Beam-focusing techniques offer new opportunities for wireless communication system design, including interference mitigation, capacity enhancement, and improved accessibility \cite{Cui2023}. Nonetheless, these techniques also pose substantial challenges to signal processing, demanding a reevaluation of channel modeling and communication techniques to accommodate the spherical wavefront. Consequently, existing wireless communication models and schemes tailored for far-field operations may become inapplicable. This requires a thorough exploration of the properties, potential benefits, and design challenges that arise when high-frequency XL-MIMO systems operate in both the far-field and near-field regions.

In light of this, several recent studies have shifted their focus to hybrid-field communication scenarios in XL-MIMO systems, where both the far-field and near-field regions coexist. Most of the research related to hybrid-field communication has concentrated on channel estimation in environments where distinct scatterers are present in both the far-field and near-field regions \cite{Wei2022, Hu2023, Yang2023}. By taking advantage of the sparsity observed in hybrid-field channels due to high-frequency bands, channel estimation schemes based on orthogonal matching pursuit (OMP) have been proposed \cite{Wei2022, Hu2023, Yang2023}. The authors in \cite{Wei2022, Hu2023} assumed that the proportion of far-field and near-field channel components can be known in advance. They further proposed an OMP-based channel estimation algorithm, which sequentially estimates the far-field channel components first and then estimates the near-field channel components. However, the approaches proposed in \cite{Wei2022, Hu2023} have a practical limitation since they necessitate having specific prior information regarding the proportion of far-field and near-field channel components, which is generally unavailable in reality. On the other hand, in \cite{Yang2023}, a channel estimation algorithm was introduced that does not rely on prior knowledge of the proportion of far-field and near-field channel components. However, it has its limitations -- it is susceptible to error propagation because it estimates the two channel components separately and sequentially, rather than jointly estimating them. 

While there have been several attempts at hybrid-field channel estimation, as aforementioned, the existing studies in this domain have limitations that fail to fully capture the characteristics of hybrid-field channels. Additionally, since compressive sensing (CS) techniques are employed to exploit channel sparsity for hybrid-field channel estimation, it is crucial to appropriately design the sensing matrix of the CS system to ensure reliable sparse recovery. However, none of the existing studies that address the hybrid-field channel estimation with CS techniques provided solutions for designing pilot signals, which is a crucial component in constructing the sensing matrix. These significant challenges have not yet been addressed in the existing literature.

To address the challenges in hybrid-field channel estimation for XL-MIMO systems, we devise two algorithms: i) a pilot signal design based on the alternating direction method of multipliers (ADMM) and ii) a two-stage sequential channel estimation algorithm involving line-of-sight (LoS) channel estimation and hybrid-field scattering channel estimation. These algorithms are tailored to the hybrid-field channel characteristics. Specifically, exploiting the inherent sparsity of the hybrid-field scattering channel due to high-frequency bands, we formulate the hybrid-field scattering channel estimation as a CS problem. To enhance the accuracy of sparse channel recovery, we first tackle the non-convex pilot signal design problem by minimizing the mutual coherence of the sensing matrix in the CS problem using the ADMM framework. Building on the acquired pilot signal, we then sequentially conduct LoS channel estimation and hybrid-field scattering channel estimation. To the best of our knowledge, this work is the first of its kind that investigates pilot signal and channel estimation co-design for hybrid-field communications. The contributions of this paper are summarized as follows.

\begin{itemize}
    \item We introduce a novel co-design of pilot signal and channel estimator tailored for hybrid-field communications. By incorporating the LoS channel component and the hybrid-field scattering channel components, we establish the hybrid-field channel estimation problem (Section \ref{sec:Preliminaries}). We then present the pilot signal design and the two-stage channel estimation algorithm, intending to initially estimate the LoS channel component and subsequently jointly estimate both the far-field and near-field channel components (Section \ref{sec:PilotDesign} and \ref{sec:ChannelEstimation}).
    \item We formulate a problem for obtaining an optimal pilot signal that minimizes mutual coherence for hybrid-field communications (Section \ref{subsec:MutualCoherenceProblem}). Moreover, we develop a method for solving this mutual coherence minimization problem, which can be challenging to solve due to non-convex constraints, by employing the ADMM framework (Section \ref{subsec:ADMMPilotDesign}). To tackle this challenge, we break down the ADMM-based pilot signal design algorithm into three steps and utilize the most suitable methods for each step.
    \item We develop a method for recovering the sparse hybrid-field channel. Initially, we estimate the LoS channel component by using a gradient descent method. Subsequently, we estimate the hybrid-field scattering channel by leveraging the Bayesian matching pursuit (BMP) method (Section \ref{sec:ChannelEstimation}). In this context, we present the BMP-based channel estimation algorithm both with and without prior channel knowledge for the hybrid-field scattering channel.
    \item The simulation results validate that the proposed co-design of the pilot signal and channel estimator yields superior performance in recovering sparse channels compared to conventional hybrid-field channel estimation algorithms (Section \ref{sec:NumericalResults}).
\end{itemize}

The remainder of this paper is organized as follows. In Section \ref{sec:Preliminaries}, we offer the preliminaries including the system model, the hybrid-field channel representation, and the problem statement. In Section \ref{sec:PilotDesign}, we elaborate on the proposed pilot signal design algorithm based on the ADMM framework. We delve into the BMP-based hybrid-field channel estimation algorithms for both scenarios, with and without prior channel knowledge in Section \ref{sec:ChannelEstimation}. Section \ref{sec:NumericalResults} provides simulation results, followed by the conclusion in Section \ref{sec:Conclusion}.

\indent \emph{Notation}: The following notations are used throughout the paper. Boldface uppercase, boldface lowercase, and normal face lowercase letters denote matrices, vectors, and scalars, respectively; $\mbX^\top$ and $\mbX^H$ denote the transpose and the conjugate (Hermitian) transpose of $\mbX$, respectively; $\mbX^\dagger$ denotes $(\mbX^H \mbX)^{-1} \mbX^H$; The \textit{p}-norm of vector $\mbx$ is denoted by $\lVert\mbx\rVert_p$ (If $p=2$, the norm is denoted by $\lVert\mbx\rVert$ without the subscript) and the Frobenius norm of matrix $\mbX$ is denoted by $\lVert\mbX\rVert_\mathrm{F}$; A typical positive integer set $\{1,2,\dots,N\}$ is represented by $\{1:N\}$; $\mathrm{ddiag}(\mbX)$ constructs diagonal matrices with the diagonal elements of $\mbX$.

%%%%%%%%%%%%%%%%%%%%%%%%%%%%%%%%%%%%%%%%%%%%%%%%%%%%%%%%%%%%%%%%%%%
% % % % % % % % % % % % % % % % % % % % % % % % % % % % % % % % % %
%%%%%%%%%%%%%%%%%%%%%%%%%%%%%%%%%%%%%%%%%%%%%%%%%%%%%%%%%%%%%%%%%%%
\section{System and Channel Models} \label{sec:Preliminaries}
%%%%%%%%%%%%%%%%%%%%%%%%%%%%%%%%%%%%%%%%%%%%%%%%%%%%%%%%%%%%%%%%%%%
% % % % % % % % % % % % % % % % % % % % % % % % % % % % % % % % % %
%%%%%%%%%%%%%%%%%%%%%%%%%%%%%%%%%%%%%%%%%%%%%%%%%%%%%%%%%%%%%%%%%%%

In this section, we elaborate on the preliminaries, which encompass the system model, the hybrid-field channel representation, and the problem statement considered for hybrid-field channel estimation in this paper.

\subsection{System Model}
\begin{figure}[t]
    \centering
    \includegraphics[width=0.7\columnwidth]{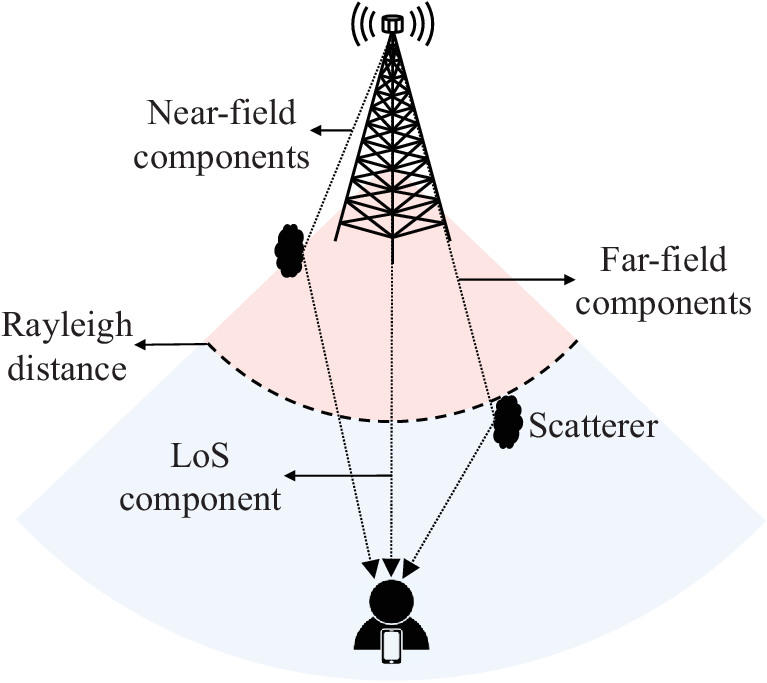}
    \caption{An illustration of a downlink XL-MIMO communication system, where the channel components are classified into two types based on the Rayleigh distance: i) far-field and ii) near-field.}
    \label{fig:SystemModel}
\end{figure}
As shown in Fig. \ref{fig:SystemModel}, we consider a downlink XL-MIMO communication system in which a base station (BS) is equipped with an $N$-element extremely large-scale ULA to communicate with a single antenna user. The BS transmits pilot sequences to the user over $M$ time slots for downlink channel estimation. The pilot signal received by the user in the $m$-th time slot is denoted as $y_m = \mbh^\top \mbx_m + w_m$, and by concatenating the received pilot signals over $M$ time slots, the received pilot signal vector $\mby \in \mathbb{C}^{M \times 1}$ can be expressed as
\begin{align} \label{ObservedVector}
    \mby = \mbX \mbh + \mbw,
\end{align}
where $\mbh \in \mathbb{C}^{N \times 1}$ denotes the channel from the BS to the user, and it is assumed that the channel remains stationary during both the pilot and data transmission phases; $\mbX = [\mbx_1,\dots,\mbx_M]^\top \in \mathbb{C}^{M \times N}$ represents the transmitted pilot signals by the BS over $M$ time slots where $\mbx_m = [x_{m,1},\dots,x_{m,N}]^\top \in \mathbb{C}^{N \times 1}$ is the pilot signal transmitted by the $N$ BS antenna elements in the $m$-th time slot; $\mbw \in \mathbb{C}^{M \times 1}$ is the noise vector, and each element of $\mbw$ follows an independent and identically distributed (i.i.d.) Gaussian distribution with zero mean and variance $\sigma_w^2$, denoted as $\mathcal{CN}(0, \sigma_w^2)$. Furthermore, we assume that the transmitted pilot signal power from the BS is $\lVert\mbx_m\rVert^2 = P_\mbx$, $\forall m \in \{1:M\}$.

The number of BS antenna elements $N$ is significantly large in an XL-MIMO system, and thus, the pilot signal vector dimension $M$ is generally smaller than $N$. This characteristic enables the use of various CS techniques that leverage the sparse channel representation, particularly in extremely high-frequency bands such as millimeter wave or sub-terahertz bands \cite{Seo2019, Kang2024}. Moreover, it allows a significant reduction in the pilot transmission duration. The following subsection will concisely explain the sparse channel representation for more efficient and effective channel estimation in far-field, near-field, and hybrid-field scenarios.

\subsection{Hybrid-Field Channel Representation}

\begin{figure}[t]
    \centering
    \includegraphics[width=0.8\columnwidth]{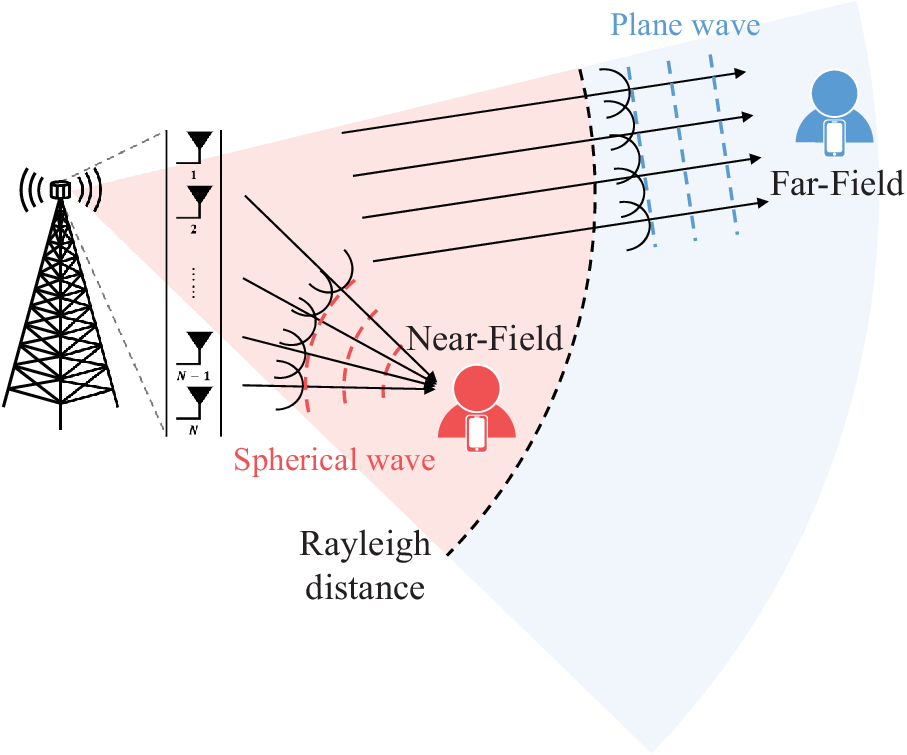}
    \caption{An illustration of the distinction between the near-field region and the far-field region, which is determined by the Rayleigh distance.}
    \label{fig:NearFieldvsFarField}
\end{figure}
It is widely recognized that when antennas radiate electromagnetic waves in a wireless medium, the waves propagate in the form of a spherical wavefront. Thus, as illustrated in Fig. \ref{fig:NearFieldvsFarField}, an electromagnetic radiation field can be divided into distinctive far-field and near-field regions based on the Rayleigh distance, denoted as $Z = \frac{2D^2}{\lambda}$, where $D$ and $\lambda$ represent the array aperture size and the carrier wavelength, respectively. In traditional wireless communication, where the communication distance is significantly greater than the Rayleigh distance, the wavefront can be accurately approximated as a plane wave. However, such an approximation is no longer valid in XL-MIMO systems, which are anticipated to employ extremely large antenna arrays and operate in extremely high-frequency bands. For instance, with an antenna aperture with a diameter of 0.5 meters at 28 GHz, the Rayleigh distance is approximately 47 meters \cite{Zhang2023}. Consequently, considering the typical coverage area in wireless communications \cite{Andrews2014}, it is highly likely that both the far-field and near-field regions coexist in an XL-MIMO system.

For the user located in the far-field region as depicted in Fig. \ref{fig:SystemModel}, we consider a hybrid-field channel model where both a direct path, referred to as an LoS channel component, and scattered paths, i.e., far-field and near-field channel components, coexist. When the distance between the BS and the scatterer exceeds the Rayleigh distance, the path components associated with the scatterer are considered to be in the far-field region. On the other hand, when the distance between the BS and the scatterer is shorter than the Rayleigh distance, the path components related to the scatterer belong to the near-field region. It is worth noting that path loss is severe in high-frequency bands, such as millimeter-wave or sub-terahertz bands. Given the significant propagation distance of the path by the scatterer far from the LoS channel component, the signal transmitted by the BS along that path faces challenges in reaching the user due to the high path loss. Therefore, the propagation distance of the scattered paths received by the user will not be significantly different from the LoS path, and the signals through the LoS path and the scattered paths may arrive within the same symbol. Consequently, a channel model is essential to accommodate the hybrid-field scenario, where the LoS channel component and the two types of scattered paths, i.e. the far-field and near-field channel components, coexist.

\subsubsection{LoS Channel Component}
To begin, we model the LoS channel component based on the geometric free-space LoS propagation assumption \cite{Sherman1962}. Consequently, the LoS channel component can be expressed as
\begin{align} \label{eq:LoSChannel}
    \left[\frac{g_\mathrm{LoS}}{r^{(n)}} e^{-j 2 \pi r^{(n)}/\lambda}\right]_{N \times 1},
\end{align}
where $r^{(n)}$ represents the distance between the $n$-th BS antenna element and the user; $g_\mathrm{LoS}$ denotes the deterministic channel gain of the LoS channel component. Furthermore, $r^{(n)}$ can be expressed as $\sqrt{r^2 + d^2 (n-1)^2 -2d r (n-1) \sin \varphi}$, where $r$ is the distance between the reference antenna element, i.e., first antenna element, of the BS and the user, and $\varphi$ denotes the angle of departure (AoD) of the signal directed to the user.

\subsubsection{Far-Field Channel Components} 
In the scattered paths, the far-field channel components are modeled based on the free-space path loss model under the far-field plane wave assumption \cite{Akdeniz2014}, which is expressed as 
\begin{align} \label{eq:FFChannel}
    \mbh^\mathrm{far} = \sqrt{\frac{N}{L_f}}\sum_{l=1}^{L_f} \frac{g_l}{r_l} \mbf^{\sfA(\theta_l)},
\end{align}
where $L_f$ represents the number of path components between the BS and the scatterers located in the far-field region; $r_l$ represents the distance between the reference antenna element of the BS and the $l$-th scatterer; $g_l$ denotes the small scale channel gain of the $l$-th path. In addition, the far-field array response vector $\mbf^{\sfA(\theta_l)}$ is given by
\begin{align} \label{eq:FFArrayVector}
    \mbf^{\sfA(\theta_l)} = \frac{1}{\sqrt{N}}[1, e^{j \pi \theta_l},\dots, e^{j (N-1) \pi \theta_l}]^H.
\end{align}
Here, $\theta_l = 2 \frac{d}{\lambda} \sin \varphi_l$ where $d$ is the antenna spacing of the array aperture, typically set as $\lambda/2$ to avoid coupling effects between the antennas. Additionally, $\varphi_l \in [-\pi/2, \pi/2]$ denotes the physical propagation angle of the $l$-th path. Thus, $\theta_l$ falls within the range $[-1, 1]$.

To exploit the channel sparsity in channel estimation, we can express the non-sparse far-field channel $\mbh^\mathrm{far}$ concerning a sparse angular domain channel representation as
\begin{align} 
    \mbh^\mathrm{far} = \mbFA \mbhA,
\end{align}
where $\mbhA = [h^\sfA_1,\dots,h^\sfA_N]^\top \in \mathbb{C}^{N \times 1}$ represents the angular domain channel; $\mbFA = [\mbf^{\sfA(\theta_1)},\dots,\mbf^{\sfA(\theta_N)}] \in \mathbb{C}^{N \times N}$ denotes a spatial discrete Fourier transform (DFT) matrix and is composed of $N$ orthogonal array response vectors uniformly sampled from the angular domain covered by the BS; $h^\sfA_n$ and $\mbf^{\sfA(\theta_n)}$ are the channel coefficient and the array response vector corresponding to $\theta_n$ in the angular domain, respectively, where $\theta_n = \frac{2n-N-1}{N}$, $\forall n \in \{1:N\}$. Due to the limited number of scatterers and the severe path loss experienced in high-frequency bands, the angular domain channel $\mbhA$ typically exhibits sparsity. 

\begin{figure}[t]
    \centering
    \includegraphics[width=0.9\columnwidth]{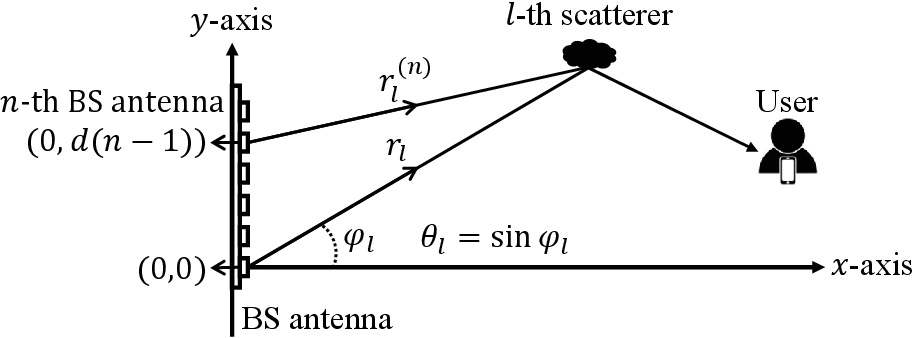}
    \caption{A schematic diagram of the near-field channel model}
    \label{fig:NearfieldCoordinate}
\end{figure}

\subsubsection{Near-Field Channel Components} \label{subsubsec:NearField}
Similar to the far-field channel, the near-field channel components are modeled under the near-field spherical wave assumption\cite{Sherman1962}. As illustrated in Fig. \ref{fig:NearfieldCoordinate}, the near-field channel can be expressed as
\begin{align} \label{eq:NFChannel}
    \mbh^\mathrm{near} = \sqrt{\frac{N}{L_n}} \sum_{l=1}^{L_n} \frac{g_l}{r_l} \mbf^{\sfP(\theta_l, r_l)},
\end{align}
and the near-field array response vector is given by
\begin{align} \label{eq:NFArrayVector}
    \mbf^{\sfP(\theta_l, r_l)} = \frac{1}{\sqrt{N}}\left[e^{-j \frac{2 \pi}{\lambda}(r_l^{(1)} - r_l)},\dots, e^{-j \frac{2 \pi}{\lambda}(r_l^{(N)} - r_l)}\right]^H,
\end{align}
where $r_l^{(n)} = \sqrt{r_l^2+d^2(n-1)^2 -2d r_l (n-1) \theta_l}$ denotes the distance between the $n$-th BS antenna element and the $l$-th scatterer.

Unlike the far-field channel, the near-field channel cannot be directly represented as a sparse angular domain channel based on the DFT matrix. This limitation arises due to the energy spread effect of the spherical wave in the near-field channel, which means that the energy of one near-field path component is not concentrated in a single direction but instead spreads in multiple directions. Therefore, when dealing with the near-field channels, both the impact of the angle and the distance must be taken into consideration. Therefore, in order to harness the channel sparsity in the near-field channel, we rephrase \eqref{eq:NFChannel} with a sparse polar domain representation as
\begin{align}
    \mbh^\mathrm{near} = \mbFP \mbhP,
\end{align}
where $\mbhP \in \mathbb{C}^{NQ \times 1}$ and $\mbFP \in \mathbb{C}^{N \times NQ}$ denote the polar domain channel and transform matrix, respectively. Note that it is assumed that the polar domain consists of uniformly sampled $N$ angles, and each angle within the polar domain is further divided into $Q$ discrete distance sections. Specifically, as outlined in \cite{Cui2022}, distance sections in the polar domain can be sampled based on the distance ring, which aids in obtaining lower column coherence for $\mbFP$. Therefore, the polar domain transform matrix $\mbFP$ can be written as
\begin{align}
    \mbFP \!=\! \left[\mbf^{\sfP(\theta_1, r_{1,1})}\!,\!...,\!\mbf^{\sfP(\theta_1, r_{1,Q})}\!,\!...,\!\mbf^{\sfP(\theta_N, r_{N,1})}\!,\!...,\!\mbf^{\sfP(\theta_N, r_{N,Q})}\right]\!,
\end{align}
where each column of $\mbFP$ is a near-field array response vector with sampled angle $\theta_n$ and sampled distance $r_{n,q}$ for $n \in \{1:N\}$ and $q \in \{1:Q\}$. As a result, even if the near-field channel is not sparse in the angular domain, it becomes sparse in the polar domain through the transformation using $\mbFP$.

\subsubsection{Hybrid-Field Channel Model}
By taking into account the LoS and the scattering channel components, we can formulate a hybrid-field channel model for XL-MIMO in terms of a sparse mixed angular and polar domain representation as
\begin{align} \label{eq:HybridChannelSparse}
    \mbh \!=& \sqrt{\frac{1}{L+1}} \left[\frac{g_\mathrm{LoS}}{r^{(n)}} e^{-j 2 \pi r^{(n)}/\lambda}\right]_{N \times 1} \nonumber \\&+ \!\sqrt{\frac{N}{L+1}} \!\left( \sum_{l_f=1}^{L_f} \frac{g_{l_f}}{r_{l_f}} \mbf^{\sfA(\theta_{l_f})} \!+\! \sum_{l_n=1}^{L_n} \frac{g_{l_n}}{r_{l_n}} \mbf^{\sfP(\theta_{l_n}, r_{l_n})} \!\right)\\
    =& \mbh^\mathrm{LoS} + \mbF \mbhAP,
\end{align}
where $L \!=\! L_f \!+\! L_n$ denotes the total number of scattered path components between the BS and all scatterers located within the coverage area of the BS; $N' \!=\! N\!+\!NQ$ for simplicity; $\mbF \!=\! \sqrt{\frac{N}{L+1}}[\mbFA, \mbFP] \in \mathbb{C}^{N \times N'}$ denotes the concatenated angular and polar domain transform matrix; $\mbhAP = [\mbhA; \mbhP] \in \mathbb{C}^{N' \times 1}$ represents the concatenated angular and polar domain channel. It is worth noting that the mixed angular and polar domain channel representation is constructed by simply concatenating the angular and polar domain channel representations, resulting in a sparse vector $\mbhAP$, such that $L \ll N'$.

\subsection{Problem Statement}
Based on the LoS path and the scattered paths, the received pilot signal vector \eqref{ObservedVector} can be reformulated as
\begin{align} \label{eq:CSProblem}
    \mby = \mbX \mbh^\mathrm{LoS} + \mbPsi \mbhAP + \mbw,
\end{align}
where $\mbPsi = \mbX \mbF \in \mathbb{C}^{M \times N'}$. In this paper, our ultimate goal is to estimate $\mbh$ from the received pilot signal vector $\mby$. We first target the recovery of $\mbh^\mathrm{LoS}$ from $\mby$ and then reconstruct $\mbhAP$ from the residual vector $\mby - \mbX \mbh^\mathrm{LoS}$. Although the problem of recovering $\mbhAP$ poses an underdetermined system challenge, because $\mbhAP$ is a sparse vector, our approach involves utilizing CS techniques to recover $\mbhAP$ from $\mby - \mbX \mbh^\mathrm{LoS}$.

However, to achieve this, there is a preliminary challenge we need to address -- ensuring the effective operation of CS techniques through the well-designed sensing matrix $\mbPsi$. Strictly speaking, since $\mbF$ is fixed, the task involves designing the pilot signal $\mbX$ in a manner that enables the sensing matrix to function properly within CS. In the literature, one of the well-known sufficient conditions for reliable sparse recovery is the restricted isometry property (RIP). It guarantees sufficient reconstruction of the sparse signal for a CS system where the stochastic sensing matrix meets the RIP. Nonetheless, when a sensing matrix is deterministic, verifying whether it satisfies the RIP is computationally infeasible. As indicated in \cite{Foucart2013}, the RIP implies that minimizing the mutual coherence between the columns of the sensing matrix is crucial for achieving reliable recovery performance. Consequently, mutual coherence is considered a more intuitive and practical measure than the RIP for assessing whether the sensing matrix ensures reliable sparse recovery, as noted in \cite{Qi2015}. Therefore, we adopt mutual coherence as the guiding principle for the pilot signal design.

Furthermore, a channel estimation algorithm suitable for the hybrid-field channel needs to be designed. We devise a two-stage channel estimation algorithm that sequentially estimates the LoS channel component and the hybrid-field scattering channel components. Firstly, a channel estimation algorithm for the LoS channel component is necessary. Additionally, a channel estimation algorithm tailored for the hybrid-field scattering channel should be paired with a well-designed sensing matrix. Unlike the existing sequential hybrid-field scattering channel estimation algorithms, which do not jointly consider far-field and near-field channel components, we aim to develop a channel estimation method that sufficiently captures the characteristics of the hybrid-field scattering channel. Consequently, we intend to address two key problems throughout this paper: the design of pilot signal and channel estimator for hybrid-field communications, to achieve more reliable sparse channel recovery.

%%%%%%%%%%%%%%%%%%%%%%%%%%%%%%%%%%%%%%%%%%%%%%%%%%%%%%%%%%%%%%%%%%%
% % % % % % % % % % % % % % % % % % % % % % % % % % % % % % % % % %
%%%%%%%%%%%%%%%%%%%%%%%%%%%%%%%%%%%%%%%%%%%%%%%%%%%%%%%%%%%%%%%%%%%
\section{Pilot Signal Design for Hybrid-Field Channel} \label{sec:PilotDesign}
%%%%%%%%%%%%%%%%%%%%%%%%%%%%%%%%%%%%%%%%%%%%%%%%%%%%%%%%%%%%%%%%%%%
% % % % % % % % % % % % % % % % % % % % % % % % % % % % % % % % % %
%%%%%%%%%%%%%%%%%%%%%%%%%%%%%%%%%%%%%%%%%%%%%%%%%%%%%%%%%%%%%%%%%%%
This section delves into the design of the sensing matrix through pilot signal design, which was previously mentioned as the first challenge. In particular, we seek to find a pilot signal matrix $\mbX$ that minimizes mutual coherence using the alternating direction method of multiplier (ADMM).

\subsection{Mutual Coherence Minimization Problem} \label{subsec:MutualCoherenceProblem}
The mutual coherence of a matrix is defined as the maximum absolute correlation between any two distinct columns of the matrix \cite{Foucart2013}. Hence, the mutual coherence $C(\mbPsi)$ of the sensing matrix $\mbPsi$ can be expressed as
\begin{align} \label{eq:MutualCoherence}
    C(\mbPsi) = \max_{1 \leq u < v \leq N'} \; \frac{\lvert \mbPsi_u^H \mbPsi_v \rvert}{\lVert \mbPsi_u \rVert_2 \lVert \mbPsi_v \rVert_2},
\end{align}
where $\mbPsi_u$ and $\mbPsi_v$ respectively represent the $u$-th and the $v$-th column vector of $\mbPsi$. For the sake of simplicity, let $\psi_{u,v} = \frac{\mbPsi_u^H \mbPsi_v}{\lVert \mbPsi_u \rVert_2 \lVert \mbPsi_v \rVert_2}$ hereafter, and denote by $\mbpsi = (\psi_{1,2},\dots,\psi_{1,N'},\psi_{2,3},\dots,\psi_{N'-1,N'})$ the mutual coherence vector, which is comprised of mutual coherence values for all combinations of two distinct columns in $\mbPsi$. Recall that the sensing matrix can be expressed as $\mbPsi = \mbX\mbF$, while the transform matrix $\mbF$ is predetermined. Thus, the pilot signal matrix $\mbX$ should be carefully designed to minimize $C(\mbPsi)$. Thus, subject to the constraints on the transmit power of the pilot signal from the BS, we can formulate the mutual coherence minimization problem $\mathcal{P}_1$ to obtain the optimal pilot signal $\mbX^*$ that minimizes $C(\mbPsi)$ as follows:
\begin{subequations}
\begin{align}
    \mathcal{P}_1 : \min_{\mbX} \quad & \max_{1 \leq u < v \leq N'} \; \lvert \psi_{u,v} \rvert \label{eq:P1Objective}\\
    \textrm{s.t.} \quad & \lVert\mbx_m\rVert^2 = P_\mbx,\ \forall m \in \{1:M\}. \label{eq:P1Const1}
\end{align}
\end{subequations}
However, solving $\mathcal{P}_1$ is challenging because it is non-convex due to the constant modulus constraints \eqref{eq:P1Const1}.

In general, optimization techniques on Riemannian manifolds are typically employed to address constant modulus constraints. However, since conventional optimization techniques on Riemannian manifolds rely on smooth functions, and non-smooth optimization techniques do not ensure convergence on Riemannian manifolds, addressing the non-smooth optimization problem on Riemannian manifolds proves to be challenging. Since the objective function in \eqref{eq:P1Objective} is non-smooth due to the presence of the maximum operator, well-known optimization techniques on Riemannian manifolds, such as gradient descent on Riemannian manifolds, cannot be directly applied to $\mathcal{P}_1$. In this case, one might consider updating $\mbX$ alternatively by solving the gradient descent subproblem on the Riemannian manifolds for given $u$ and $v$ in all combinations of $u$ and $v$ through relaxation of the maximum operator. However, considering that the number of combinations of $u$ and $v$ is on the order of $\mathcal{O}(N'^2)$, hybrid XL-MIMO systems would necessitate alternative updates for a significant number of $\mathcal{O}(N'^2)$ subproblems, making it a computationally expensive process. Consequently, to overcome these challenges, a method is needed to solve the optimization problem by separating the two problematic factors. Hence, we adopt the ADMM method, which can decompose the non-smooth objective function and the constant modulus constraints \cite{Boyd2011}. It is worth noting that, while an alternative update algorithm by the gradient descent subproblems on the Riemannian manifolds incurs a computational complexity of $\mathcal{O}(M N N'^2 + N^3 N'^2)$ per iteration, ADMM requires a lower cost of $\mathcal{O}(M N N'^2 + N^3)$.

The key idea of ADMM is to introduce an auxiliary vector $\mbxi = \left(\xi_1, \dots,\xi_{N' \choose 2}\right)$ to redefine the objective function. ADMM provides an equivalent optimization by augmenting the objective function with constraints, introducing the auxiliary variables $\mbxi$ for the constraints $\mbpsi$, and enforcing $\mbxi$ to approach $\mbpsi$. Then, $\mathcal{P}_1$ can be recast as follows:
\begin{subequations}
\begin{align}
    \mathcal{P}_2 : \min_{\mbxi, \mbX} \quad & \lVert \mbxi \rVert_\infty\\
    \textrm{s.t.} \quad & \mbxi = \mbpsi \label{eq:P2Const1}\\
    &\lVert\mbx_m\rVert^2 = P_\mbx, \ \forall m \in \{1:M\}, \label{eq:P2Const2}
\end{align}
\end{subequations}
where $\lVert \mbxi \rVert_\infty = \max_{1 \leq i \leq {N' \choose 2}} \; \lvert \xi_{i} \rvert$.

\subsection{Pilot Signal Design} \label{subsec:ADMMPilotDesign}
Following the principles of the ADMM framework, we can define the augmented Lagrangian function for $\mathcal{P}_2$, which includes the objective function, the dual function, and a penalty term on the objective function:
\begin{align}
    L(\mbxi, \mbX, \mblamb) =& \lVert \mbxi \rVert_\infty \!+\! \mathbb{I}(\mbX) \!+\! \langle \mblamb, \mbxi - \mbpsi \rangle \!+\! \frac{\rho}{2}\lVert \mbxi - \mbpsi \rVert_2^2, \label{eq:LagrangianFunction}
\end{align}
where $\mblamb$ is the Lagrange multiplier; $\rho > 0$ is the penalty parameter; $\langle \mbx,\mby \rangle \triangleq \Re \{ \mby^H \mbx\}$ denotes a vector inner product operation. Furthermore, $\mathbb{I}(\mbX)$ is an indicator function representing the constant modulus constraints \eqref{eq:P2Const2}:
\begin{align}
    \mathbb{I}(\mbX) = 
    \bigg\{ \begin{array}{ll}
    0, & \mbox{if $\lVert\mbx_m\rVert^2 = P_\mbx,\ \forall m \in \{1:M\}$}\\
    \infty, & \mbox{otherwise}.\end{array}
\end{align}
Based on $L(\mbxi, \mbX, \mblamb)$ in the ADMM framework, we can solve the problem by alternatively updating $\mbxi$, $\mbX$, and $\mblamb$. In consequence, the ADMM algorithm consists of three steps, and the ADMM iteration steps are as follows:
\begin{align}
    \mbxi^{(l)} &= \argmin_{\mbxi} L(\mbxi, \mbX^{(l-1)}, \mblamb^{(l-1)}), \label{eq:update_xi}\\
    \mbX^{(l)} &= \argmin_{\mbX} L(\mbxi^{(l)}, \mbX, \mblamb^{(l-1)}), \label{eq:update_X}\\
    \mblamb^{(l)} &= \mblamb^{(l-1)} + \rho(\mbxi^{(l)}-\mbpsi^{(l)}). \label{eq:update_lamb}
\end{align}
The ADMM should update the primal variables $\mbxi, \mbX$, followed by updating dual variable $\mblamb$. However, in the update process of $\mbxi$ and $\mbX$, they can be updated in an arbitrary order, and even so, convergent results can be achieved \cite{Wang2019}. The detailed process of the ADMM algorithm is described below.
\begin{enumerate}
    \item[(a)] \textbf{Update $\mbxi$}: First, the objective function for updating $\mbxi$ is
\begin{equation}
    \begin{aligned}
    \min_{\mbxi} \; \lVert \mbxi \rVert_\infty + \langle \mblamb, \; \mbxi - \mbpsi \rangle + \frac{\rho}{2}\lVert \mbxi - \mbpsi \rVert_2^2. \label{eq:ObjectiveXi}
    \end{aligned}
\end{equation}
In \eqref{eq:ObjectiveXi}, we can modify $\langle \mblamb, \; \mbxi - \mbpsi \rangle + \frac{\rho}{2}\lVert \mbxi - \mbpsi \rVert_2^2$ as
\begin{align}
    &\langle \mblamb, \; \mbxi - \mbpsi \rangle + \frac{\rho}{2}\lVert \mbxi - \mbpsi \rVert_2^2\nonumber\\
    &= \Re \{ (\mbxi - \mbpsi)^H \mblamb\} + \frac{\rho}{2}\lVert \mbxi - \mbpsi \rVert_2^2\\
    &= \frac{1}{2} \left( \mblamb^H (\mbxi - \mbpsi) + (\mbxi - \mbpsi)^H \mblamb \right) + \frac{\rho}{2}\lVert \mbxi - \mbpsi \rVert_2^2 \label{eq:ObjectiveXiEq1}\\
    &= \frac{\rho}{2} \left\lVert \mbxi - \mbpsi + \frac{\mblamb}{\rho} \right\rVert_2^2 + C,
\end{align}
where \eqref{eq:ObjectiveXiEq1} is derived from $\langle \mblamb, \; \mbxi - \mbpsi \rangle = \Re \{ (\mbxi - \mbpsi)^H \mblamb\} = \frac{1}{2} (\mblamb^H (\mbxi - \mbpsi) + (\mbxi - \mbpsi)^H \mblamb )$; The constant $C = -\frac{\mblamb^H \mblamb}{2 \rho}$ is irrelevant to $\mbxi$. Therefore, the subproblem for updating $\mbxi$ corresponding to \eqref{eq:update_xi} is formulated as
\begin{equation}
    \begin{aligned}
    \mathcal{P}_{\mbxi} : \min_{\mbxi} \; \lVert \mbxi \rVert_\infty + \frac{\rho}{2} \left\lVert \mbxi - \mbpsi^{(l-1)} + \frac{\mblamb^{(l-1)}}{\rho} \right\rVert_2^2. \label{eq:Problemxi}
    \end{aligned}
\end{equation}
It is worth noting that $\mathcal{P}_{\mbxi}$ is an unconstrained optimization problem that includes both a least-squares term and an infinity norm term. This problem falls under the category of convex optimization and can be equivalently transformed into a convex semidefinite problem (SDP). Consequently, by utilizing convex optimization tools such as SDPT3 \cite{Toh1999}, we can obtain the optimal solution $\mbxi^*$ that minimizes the objective function in \eqref{eq:Problemxi}.

    \item[(b)] \textbf{Update $\mbX$}: Similar to updating $\mbxi$, the subproblem for updating $\mbX$ corresponding to \eqref{eq:update_X} can be formulated as
\begin{subequations}\label{eq:ProblemX}
    \begin{align}
    \mathcal{P}_{\mbX} : \min_{\mbX} \; & \left\lVert \mbpsi - \mbxi^{(l)} - \frac{\mblamb^{(l-1)}}{\rho} \right\rVert_2^2\\
    \textrm{s.t.} \; & \lVert\mbx_m\rVert^2 = P_\mbx \text{ for } \forall m \in \{1:M\}. \label{eq:ProblemXConst}
    \end{align}
\end{subequations}
However, due to the non-convex constant modulus constraints \eqref{eq:ProblemXConst}, solving $\mathcal{P}_{\mbX}$ is challenging. To address this difficulty, we can harness the smooth Riemannian manifold structure inherent in the constant modulus constraints, which are on the complex oblique manifold $\mathcal{M}_\mbX \!=\! \{\mbX \!\in\! \mathbb{C}^{M \times N}\!:\! \mathrm{ddiag}(\mbX^H \mbX) \!=\! P_\mbx\mathbf{I}\}$ \cite{Absil2008}. This approach allows us to transform the non-convex problem $\mathcal{P}_{\mbX}$ into an unconstrained optimization problem on the Riemannian manifold. The core concept behind Riemannian manifold optimization is to devise a gradient descent algorithm specifically tailored to the Riemannian manifold. By employing Riemannian stochastic gradient descent (SGD) method designed for Riemannian optimization problems, we can find the optimal pilot signal $\mbX^*$, which converges towards a Riemannian zero-gradient point \cite{Absil2008}.

    \item[(c)] \textbf{Update $\mblamb$}: Finally, after updating $\mbxi$ and $\mbX$, we update $\mblamb$ using the dual ascent method as shown in \eqref{eq:update_lamb}.
\end{enumerate}

As a result, we employ ADMM to address and decompose a mutual coherence minimization problem involving two challenging factors. Furthermore, within the ADMM framework, most research utilizing ADMM employs a fixed penalty parameter $\rho$. However, using a fixed $\rho$ often results in slower convergence due to the imbalance between the objective and the residual. As evident from \eqref{eq:LagrangianFunction}, when $\rho$ is excessively large, the augmented Lagrangian function becomes dominated by the penalty term. This leads to overemphasizing the primal and dual residuals, yielding less effective solutions. Conversely, if $\rho$ is too small, an excessive focus on the objective may fail to ensure satisfying primal and dual feasibility conditions. Therefore, we advocate for employing a variable $\rho$, which gradually increases from 0.05 to 2.0. This choice aims to achieve both faster convergence and robust performance. Through this iterative process, we ultimately arrive at the optimal pilot signal $\mbX^*$, which minimizes $C(\mbPsi)$, thus enhancing sparse channel recovery performance \cite{Boyd2011}. The mutual coherence depending on $\rho$ will be covered in the next subsection.

\subsection{Mutual Coherence for Various Pilot Signals}

\begin{table}[t]
    \caption{Comparison of Mutual Coherence for Various Pilot Signals}\label{tab:MutualCoherence}
    \small
    \centering
    \begin{tabular}{l|c}
        \toprule\toprule
        \textbf{Pilot Signal} & \textbf{Mutual Coherence} \\ 
        \midrule
        Random binary sequence & 0.7996\\
        Unimodular sequence with random phase & 0.8015\\
        Zadoff-Chu sequence & 0.9574\\
        \textbf{Proposed ADMM-based sequence} & \textbf{0.4966}\\
        \bottomrule\bottomrule
    \end{tabular}
\end{table}

Table \ref{tab:MutualCoherence} provides a summary of mutual coherence, as defined in \eqref{eq:MutualCoherence}, for various pilot signals in a scenario with a pilot length of $M = 32$, $N = 128$ BS antenna elements, which is the same environment covered in Section \ref{sec:NumericalResults}. The table presents mutual coherence values for $\mbPsi = \mbX \mbF$ with different types of $\mbX$. The mutual coherence values for well-known conventional pilot signals, such as random binary sequence, unimodular sequence with random phase, and Zadoff-Chu sequence, are presented.

It is observed that the common random sequences, including the random binary sequence and unimodular sequence with random phase, exhibit an average mutual coherence of approximately 0.8 over independent 1000 trials. In contrast, the Zadoff-Chu sequence demonstrates considerably higher mutual coherence of around 0.95. This is because the Zadoff-Chu sequence is designed to minimize correlations within pilot signals. As a result, it may not achieve the desired performance when used in conjunction with the concatenated angular and polar domain transform matrix $\mbF$ for hybrid-field channel estimation. However, our proposed ADMM-based pilot signal design algorithm aims to minimize the mutual coherence of $\mbPsi$, thereby significantly improving mutual coherence. Furthermore, we examine the mutual coherence of a random Gaussian matrix of the same size as $\mbPsi$. Over independent 1000 trials, we find that the average mutual coherence of the random Gaussian matrix is 0.6435. This finding affirms the effectiveness of the proposed pilot signal design algorithm, indicating significantly good performance.

Specifically, as mentioned earlier, we confirmed that, for fixed penalty parameters of 0.1, 0.5, 1, and 2, the corresponding mutual coherence values are 0.5990, 0.5239, 0.5675, and 0.6485, respectively. Additionally, the norms of the differences between $\mbxi$ and $\mbpsi$ are 7.3988, 0.1279, 0.0825, and 0.0709, respectively. While a smaller penalty parameter results in smaller mutual coherence, it is not reasonable to conclude that the convergence state has been reached due to the relatively large norm of the difference between $\mbxi$ and $\mbpsi$. On the contrary, it can be inferred that the proposed algorithm has achieved a convergence state, as evidenced by a mutual coherence of about 0.4966 and a norm difference between $\mbxi$ and $\mbpsi$ of 0.0542. These results demonstrate that varying the penalty parameter facilitates faster convergence and more robust performance.

%%%%%%%%%%%%%%%%%%%%%%%%%%%%%%%%%%%%%%%%%%%%%%%%%%%%%%%%%%%%%%%%%%%
% % % % % % % % % % % % % % % % % % % % % % % % % % % % % % % % % %
%%%%%%%%%%%%%%%%%%%%%%%%%%%%%%%%%%%%%%%%%%%%%%%%%%%%%%%%%%%%%%%%%%%
\section{Two-Stage Hybrid-Field Channel Estimation} \label{sec:ChannelEstimation}
%%%%%%%%%%%%%%%%%%%%%%%%%%%%%%%%%%%%%%%%%%%%%%%%%%%%%%%%%%%%%%%%%%%
% % % % % % % % % % % % % % % % % % % % % % % % % % % % % % % % % %
%%%%%%%%%%%%%%%%%%%%%%%%%%%%%%%%%%%%%%%%%%%%%%%%%%%%%%%%%%%%%%%%%%%

This section explores the channel estimation algorithm, considering both the LoS channel component and the hybrid-field scattering channel components, encompassing both far-field and near-field channel components, as previously mentioned in the second challenge. We propose a two-stage hybrid-field channel estimation algorithm that includes i) LoS channel estimation and ii) hybrid-field scattering channel estimation. While the LoS channel component is determined by the distance and AoD between the BS and the user, the scattering channel components exhibit stochastic behavior depending on the scatterers. Consequently, we consider channel estimation algorithms suitable for the characteristics of each channel component and intend to sequentially estimate the LoS channel component $\mbh^\mathrm{LoS}$ and the hybrid-field scattering channel components $\mbhAP$ from $\mby$ based on considered channel estimation algorithms for each stage.

\subsection{LoS Channel Estimation}
Since the energy of the LoS channel component is typically dominant in high-frequency bands, we initially conduct LoS channel estimation from the received pilot signal vector $\mby$. In \eqref{eq:LoSChannel}, the LoS channel component depends on two parameters: the distance between the reference antenna element of the BS and the user $r$, and the AoD of the signal directed to the user $\varphi$. Therefore, with $\mbX^*$ achieved in the previous section\footnote{It is worth noting that since $\mbPsi = \mbX^* \mbF$, the mutual coherence of $\mbPsi$ includes that of $\mbX^*$. Therefore, even though $\mbX^*$ was designed to minimize the mutual coherence of $\mbPsi$ in the previous section, the mutual coherence of $\mbX^*$ is indeed minimized. We validate that the mutual coherence of $\mbX^*$ is about 0.48, demonstrating that sufficient recovery performance can be achieved even when using $\mbX^*$, designed in the previous section, for LoS channel estimation.}, we treat the LoS channel estimation problem as a parameter estimation problem concerning $r$ and $\varphi$, which can be expressed as
\begin{align} \label{eq:LoSChannelEstimationProblem}
    \min_{r, \varphi} \; & J(r, \varphi) \triangleq \left\lVert \mby - \mbX^* \mbh^\mathrm{LoS} \right\rVert.
\end{align}

We opt for the gradient descent algorithm\footnote{One common option to estimate the parameters in $\mbh^\mathrm{LoS}$ is to use the minimum mean square error (MMSE) estimate. However, obtaining a precise mathematical knowledge of $y - \mbX^* \mbh^\mathrm{LoS}$ for the feasible set of $\{r, \varphi\}$ might be challenging. In such cases, the gradient descent algorithm provides a viable alternative for the optimization.}, widely employed for solving parameter estimation problems, in LoS channel estimation \cite{Lu2023}. To initialize the values for $r$ and $\varphi$ before applying the gradient descent algorithm, we conduct on-grid coarse parameter estimation. The on-grid coarse parameter estimation aims to find initial parameters that satisfy $\eqref{eq:LoSChannelEstimationProblem}$ among all candidate parameters $\{r, \varphi\} \in \Lambda$, where $r = r_\mathrm{min}, r_\mathrm{min}+\Delta r,...,r_\mathrm{max}$ and $\varphi = \varphi_\mathrm{min}, \varphi_\mathrm{min}+\Delta \varphi,...,\varphi_\mathrm{max}$.
Here, $r_\mathrm{min}, r_\mathrm{max}, \varphi_\mathrm{min}, \varphi_\mathrm{max}$ represent lower and upper boundaries of $r$ and $\varphi$, respectively, determined by the communication coverage of the BS. Through the on-grid coarse parameter estimation, we can obtain
\begin{align}
    \hat{r}^{(0)}, \hat{\varphi}^{(0)} = \argmin_{\{r, \varphi\} \in \Lambda} \left\lVert \mby - \mbX^* \mbh^\mathrm{LoS} \right\rVert.
\end{align}

After obtaining the initial parameters $\hat{r}^{(0)}$ and $\hat{\varphi}^{(0)}$, we employ the gradient descent algorithm to obtain accurate parameter estimates for LoS channel estimation. We iteratively refine the two parameters $r$ and $\varphi$ from their initial values $\hat{r}^{(0)}$ and $\hat{\varphi}^{(0)}$. The objective function $J(r, \varphi)$ is minimized using an iterative gradient descent method, and the parameters at $t$-th iteration are updated as
\begin{align}
    \hat{r}^{(t)} &\leftarrow \hat{r}^{(t-1)} - \eta_r \nabla_r J^{(t-1)}(\hat{r}^{(t-1)}, \hat{\varphi}^{(t-1)}),\\
    \hat{\varphi}^{(t)} &\leftarrow \hat{\varphi}^{(t-1)} - \eta_\varphi \nabla_\varphi J^{(t-1)}(\hat{r}^{(t-1)}, \hat{\varphi}^{(t-1)}),
\end{align}
where $\eta_r$ and $\eta_\varphi$ represent the learning rates ensuring that $J^{(t)}(\hat{r}^{(t)}, \hat{\varphi}^{(t)}) \leq J^{(t-1)}(\hat{r}^{(t-1)}, \hat{\varphi}^{(t-1)})$. As the number of iterations increases, the accuracy of the parameter estimates improves. The gradient descent algorithm for LoS channel estimation is iterated until the difference in the objective function between the previous and current iterations is smaller than a specified threshold $\epsilon$. Therefore, the gradient descent algorithm yields the parameter estimates $\hat{r}$ and $\hat{\varphi}$ and the LoS channel component estimate $\hat{\mbh}^\mathrm{LoS}$. Consequently, by subtracting $\mbX^* \hat{\mbh}^\mathrm{LoS}$ from $\mby$, we obtain the residual vector $\bmby = \mby - \mbX^* \hat{\mbh}^\mathrm{LoS}$ for hybrid-field scattering channel estimation.

\subsection{Hybrid-Field Scattering Channel Estimation}

After estimating $\hat{\mbh}^\mathrm{LoS}$, we aim to estimate the hybrid-field scattering channel components from $\bmby$. We propose hybrid-field scattering channel estimation algorithms that leverage Bayesian matching pursuit (BMP), building upon the optimal pilot signal introduced in Section \ref{sec:PilotDesign}. While various CS algorithms like OMP, IHT, and BCS have been employed for channel estimation, Bayesian approaches are known to offer superior sparse recovery performance compared to other techniques \cite{Masood2013} and are anticipated to exploit the different distributions of the far-field and near-field channel components sufficiently. However, Bayesian approaches typically assume prior knowledge of the sparse signal, often modeling it as Gaussian. In practical scenarios, this information might be unavailable, or the sparse signal may follow an unknown distribution. Consequently, we aim to devise channel estimation algorithms for both scenarios, with and without prior statistical information about the hybrid-field scattering channel.

\subsubsection{With Prior Channel Knowledge} \label{subsec:ChannelEstimationWith}

First, we consider the scenario where prior statistical information about the hybrid-field scattering channel is available. To employ the Bayesian framework for estimating the hybrid-field scattering channel, we establish appropriate statistical assumptions considering the characteristics of the hybrid-field scattering channel. Based on experimental measurements from \cite{Russer2015}, utilizing a Gaussian distribution as the prior statistical channel model is suitable for high-frequency systems. Recall that the concatenated angular and polar domain channel $\mbhAP$ represents the small scale channel gain $g_n$ divided by the distance $r_n$ from the reference antenna element of the BS to the 
grid corresponding to the index $n$ and then to the user in the mixed angular and polar domain, i.e., $\hAP_n = g_n/r_n$ for $\forall n \in \{1:N'\}$. Furthermore, $r_n$ can be readily obtained from the estimated position of the user in the LoS channel estimation stage.

For the hybrid-field scattering channel, $\{g_n\}_{n=1}^{N'}$ are assumed to be drawn from three specific Gaussian distributions, each modeled by a sparsity pattern $s_n = \nu$ for $\nu \in \{\emptyset, \sfA, \sfP\}$, implying the support of the sparse channel. We define the set of indices with ${s_n} \in \{\sfA, \sfP\}$ among $\{1:n\}$ as the support $\cS$ of the sparse channel. When $s_n = \emptyset$, it implies that $g_n = 0$, i.e. $\{ \mu_{s_n}, \sigma_{s_n} \} = \{0, 0\}$. On the other hand, when $s_n = \sfA$ or $\sfP$, it means that $g_n$ follows the distribution of the far-field channel $\mathcal{CN}(0, \sigma_\sfA^2)$ or the near-field channel $\mathcal{CN}(0, \sigma_\sfP^2)$, respectively. Consequently, the angular domain channel gain $\{g_n\}_{n=1}^{N}$ follows either $\mathcal{CN}(0, \sigma_\sfA^2)$ or $0$, and the polar domain channel gain $\{g_n\}_{n=N+1}^{N'}$ follows either $\mathcal{CN}(0, \sigma_\sfP^2)$ or $0$. Remarkably, nonzero channel components in $\mbhAP$ imply the presence of the effective scatterer, as illustrated in Fig. \ref{fig:SystemModel}. In other words, each element of the sparse pattern signifies whether the effective scatterer exists at a location corresponding to that element. To model the effective scatterer, i.e., the sparsity of $\mbhAP$, it is assumed that $\{s_n\}_{n=1}^{N'}$ are i.i.d. random variables with $\mathbb{P}(s_n) \in \{p_\emptyset, p_{\sfA, \sfP}\}$, where $\mathbb{P}(s_n) = p_\emptyset$ and $\mathbb{P}(s_n) = p_{\sfA, \sfP}$ indicate the probability that the $n$-th element of the channel vector is zero and nonzero, respectively. Consequently, the hybrid-field scattering channel can be modeled using a Bernoulli-complex Gaussian distribution, and the value of $p_{\sfA, \sfP}$ is considerably smaller than 1, indicating the sparsity of the channel.

From the statistical model for the hybrid-field scattering channel in \eqref{eq:CSProblem}, the residual vector $\bmby$ and the sparse channel $\mbhAP$ are jointly Gaussian, given the sparsity pattern $\mbs$:
\begin{align} \label{eq:JointGaussian}
    \left[\! \begin{array}{c}
        \bmby \\
        \mbhAP
    \end{array} \!\right]\bigg\vert \; \mbs \sim \mathcal{CN}
    \left( \left[\! \begin{array}{c}
        \mathbf{0} \\
        \mathbf{0}
    \end{array} \!\right], \left[\! \begin{array}{cc}
        \mbGam(\mbs) \!&\! \mbPsi \mbK(\mbs)\\
        \mbK(\mbs) \mbPsi^H \!&\! \mbK(\mbs)
    \end{array} \!\right] \right),
\end{align}
where $\mbK(\mbs)$ is a diagonal matrix with $[\mbK(\mbs)]_{n,n} \in \{0, \sigma_\sfA^2/r_n^2, \sigma_\sfP^2/r_n^2\}$; $\mbGam(\mbs) \triangleq \mbPsi \mbK(\mbs) \mbPsi^H + \sigma_w^2 \mbI_M$. For channel estimation, the minimum mean square error (MMSE) estimate of $\mbhAP$ conditioned on $\bmby$ can be expressed as
\begin{align}
    \hmbhAP = \sum_{\mbs} \mathbb{P}(\mbs|\bmby) \mathbb{E}\left[\mbhAP \vert \mbs, \bmby \right],
\end{align}
and from \eqref{eq:JointGaussian}, we can readily obtain \cite{Poor1994}
\begin{align} \label{eq:ChannelEstimate}
    \mathbb{E}\left[\mbhAP \vert \mbs, \bmby \right] = \mbK(\mbs) \mbPsi^H \mbGam(\mbs)^{-1}\bmby.
\end{align}
However, it is computationally infeasible to compute $\mathbb{P}(\mbs|\bmby)$ for all possible combinations of $\mbs$, and obtaining $\mbGam(\mbs)^{-1}$ in $\mathbb{E}\left[\mbhAP \vert \mbs, \bmby \right]$ is challenging. Therefore, to address these challenges, we choose to utilize BMP for computationally efficient estimation of $\mbs$ and $\mbhAP$.

BMP is a greedy algorithm that sequentially identifies the nonzero channel components in each iteration, similar to the other greedy CS algorithms such as OMP. Initially, our objective is to find $\mbs$ with the largest posterior probability $\mathbb{P(\mbs \vert \bmby)}$. Using Bayes' rule, the posterior probability is expressed as
\begin{align}
    \mathbb{P}(\mbs\vert\bmby) = \frac{\mathbb{P}(\bmby\vert\mbs) \mathbb{P}(\mbs)}{\mathbb{P}(\bmby)}.
\end{align}
Since the value of $\mathbb{P}(\bmby)$ is constant for a given $\bmby$, $\mathbb{P}(\mbs\vert\bmby)$ is proportional to $\mathbb{P}(\bmby\vert\mbs) \mathbb{P}(\mbs)$. Therefore, we define $\mathbb{P}(\bmby\vert\mbs) \mathbb{P}(\mbs)$ in the log domain as the sparsity decision metric $\alpha(\mbs, \bmby)$:
\begin{align}
    \alpha(\mbs, \bmby) \triangleq& \ln \mathbb{P}(\bmby\vert\mbs) \mathbb{P}(\mbs) \label{eq:SparsityDecision}\\
    =& -\frac{1}{2}\bmby^H \mbGam(\mbs)^{-1} \bmby - \frac{1}{2}\ln \det (\mbGam(\mbs)) \nonumber \\ 
    & - \frac{M}{2} \ln(2 \pi) + \sum_{n=1}^{N'} \ln p_{{s_n}}. \label{eq:SparsityDecision2}
\end{align}
As a result, by using the sparsity decision metric $\alpha(\mbs, \bmby)$, we can iteratively obtain the optimal sparsity pattern through the BMP algorithm.

The process of finding the sparsity pattern $\mbs$ initializes from an initial state ($\mbs^{(0)} = \mathbf{0}$ and $\cS = [\;]$), and the sparsity decision metric for the initial state can be expressed as
\begin{align}
    \alpha(\mbs^{(0)}, \bmby) = -\frac{1}{2 \sigma_w^2} \lVert \bmby \rVert^2 - \frac{M}{2} \ln (2 \pi \sigma_w^2) + N' \ln p_\emptyset.
\end{align}
Then, we identify the element with the largest sparsity decision metric as the optimal element at each iteration. When the $i$-th element of the sparsity pattern is selected as the optimum at the $l$-th iteration, the sparsity pattern $\mbs^{(l)}$ is updated by incorporating this selected $i$-th element into the sparsity pattern $\mbs^{(l-1)}$ at the previous iteration, and the index $i$ is included in $\cS$. Moreover, the covariance matrix in $\alpha(\mbs, \bmby)$ is updated as 
\begin{align} \label{eq:Covariance}
    \mbGam(\mbs^{(l)}) = \mbGam(\mbs^{(l-1)}) + \sigma_{s_i}^2 \mbPsi_i \mbPsi_i^H/r_i^2.
\end{align}
By using the matrix inversion lemma, the inversion of $\mbGam(\mbs^{(l)})$ can be expressed as
\begin{align} \label{eq:CovarianceInverse}
    \mbGam(\mbs^{(l)})^{-1} \!\!=\! \mbGam(\mbs^{(l-1)})^{-1} \!\!-\! \sigma_{s_i}^2 \beta_i^{(l-1)} \mbq^{(l-1)}_i \!(\mbq_i^{(l-1)})^H\!/r_i^2,
\end{align}
where $\beta_i^{(l-1)} = (1 + \sigma_{s_i}^2 \mbPsi_i^H \mbq_i^{(l-1)}/r_i^2)^{-1}$; $\mbq_i^{(l-1)} = \mbGam(\mbs^{(l-1)})^{-1} \mbPsi_i$. From \eqref{eq:SparsityDecision2}, \eqref{eq:CovarianceInverse}, the sparsity decision metric at the $l$-th iteration can be obtained in a recursive form as \cite{Schniter2008}
\begin{align}
    \alpha(\mbs^{(l)}, \bmby) = &\alpha(\mbs^{(l-1)}, \bmby) + \frac{\sigma_{s_i}^2}{2 r_i^2} \beta_i^{(l-1)} \left\lvert \bmby^{H} \mbq_i^{(l-1)} \right\rvert^2 \nonumber\\ 
    &+ \frac{1}{2} \ln \beta_i^{(l-1)} + \ln \frac{p_{\sfA, \sfP}}{p_\emptyset}.
\end{align}
Furthermore, $\{\mbq_n^{(l)}\}_{n=1}^{N'}$ can be updated as
\begin{align}
    \mbq_n^{(l)} \!&= \!\left[ \mbGam(\mbs^{(l-1)})^{-1} \!\!-\! \sigma_{s_i}^2 \beta_{i}^{(l-1)} \mbq_{i}^{(l-1)} (\mbq_{i}^{(l-1)})^{H}\!/r_i^2 \right]\! \mbpsi_n\\
    &= \mbq_{n}^{(l-1)} - \sigma_{s_i}^2 \beta_{i}^{(l-1)} \mbq_{i}^{(l-1)} (\mbq_{i}^{(l-1)})^{H} \mbpsi_n/r_i^2.
\end{align}

Since the parameters for selecting the most probable channel element among the remaining elements can be updated recursively at each iteration, the complexity of computing $\mbGam(\mbs)^{-1}$ can be significantly reduced, enabling us to efficiently find $\mbs$. Furthermore, by utilizing \eqref{eq:ChannelEstimate}, we can obtain the estimate of $\mbhAP$ for $\mbs$. The detailed process of the proposed BMP-based hybrid-field scattering channel estimation with prior channel knowledge is summarized in Algorithm \ref{al:ChannelEstimation1}.

\begin{algorithm}[t]
\caption{BMP-based Hybrid-Field Scattering Channel Estimation with Prior Channel Knowledge} \label{al:ChannelEstimation1}
\DontPrintSemicolon
\SetKwProg{Init}{Initialize:}{}{}
\Init{$\mbs^{(0)} = \mathbf{0}, \alpha(\mbs^{(0)}, \bmby), \cS = [\;]$}{
}
\For{$l = 1:L$}{
    \For{$n \in \{1:N'\}\setminus \cS$}{
        \eIf{$l=1$}{
            $\mbq_n^{(l-1)} = \frac{1}{\sigma_w^2} \mbpsi_n$ \;
            }{
            $\mbq_n^{(l-1)} = \mbq_{n}^{(l-2)} - \sigma_{s_{n^*}}^2 \beta_{n^*}^{(l-2)} \mbq_{n^*}^{(l-2)} (\mbq_{n^*}^{(l-2)})^{H} \mbpsi_n/r_{n^*}^2$ \;
            }
        $\beta_n^{(l-1)} = \left(1+ \sigma_{s_n}^2 \mbpsi_n^H \mbq_n^{(l-1)}/r_n^2\right)^{-1}$ for $s_n \in \{ \sfA, \sfP \}$ \;
        $\alpha_n^{(l)}$ = $\alpha(\mbs^{(l-1)}, \bmby) + \frac{\sigma_{s_n}^2}{2 r_n^2} \beta_n^{(l-1)} \left\lvert \bmby^{H} \mbq_n^{(l-1)} \right\rvert^2 + \frac{1}{2} \ln \beta_n^{(l-1)} + \ln \frac{p_{\sfA, \sfP}}{p_\emptyset}$
    }
    $n^* \gets$ index with the largest $\alpha_n^{(l)}$ \;
    $\mbs^{(l)} = [\mbs^{(l-1)}, s_{n^*}]$ and $\cS \gets \cS \cup \{n^*\}$\;
    $\alpha(\mbs^{(l)}, \bmby) = \alpha_{n^*}^{(l)}$
}
$\mbQ = [\mbq_1^{(L)},\dots,\mbq_{N'}^{(L)}]$ \;
$\hmbhAP = \mbK(\mbs^{(L)}) \mbQ^H \bmby$ and $\hmbh = \mbF \hmbhAP$
\end{algorithm}

\subsubsection{Without Prior Channel Knowledge} \label{subsec:ChannelEstimationWithout}

Next, we explore a scenario where prior statistical information about the hybrid-field scattering channel is unavailable. The absence of prior statistical information hinders the definition of the sparsity pattern, allowing only the identification of the support of the sparse channel denoted as $\cS$. Additionally, computing the MMSE estimate $\mathbb{E}\left[\mbhAP \vert \cS, \bmby \right]$ becomes highly challenging. Instead, the best alternative is to use the best linear unbiased estimate, which is the ordinary least squares estimate \cite{Puntanen1989}:
\begin{align} \label{eq:ChannelEstimateWithout}
    \mathbb{E}\left[\mbhAP \vert \cS, \bmby \right] = \left( \mbPsi_\cS^H \mbPsi_\cS \right)^{-1} \mbPsi_\cS \bmby,
\end{align}
where $\mbPsi_\cS$ is a submatrix of $\mbPsi$ consisting of columns indexed by $\cS$.

Similar to the case with prior channel knowledge, we choose to leverage BMP for estimating $\cS$ and $\mbhAP$. Given $\cS$, $\mbhAP$ and $\bmby$ are Gaussian, which can be expressed as follows:
\begin{align}
    &\mbhAP \vert \cS \sim \mathcal{CN}(\mathbf{0}, \mbK(\cS)),\\
    &\bmby \vert \cS \sim \mathcal{CN}(\mathbf{0}, \mbGam_\cS).
\end{align}
Here, if $n \in \cS$, $[\mbK(\cS)]_{n,n} \in \{\sigma_\sfA^2/r_n^2, \sigma_\sfP^2/r_n^2\}$; otherwise, it is zero; $\mbK_\cS \in \mathbb{C}^{L \times L}$ denotes a diagonal matrix comprising solely nonzero diagonal elements from $\mbK(\cS)$; $\mbGam_\cS \triangleq \mbPsi_\cS \mbK_\cS \mbPsi_\cS^H + \sigma_w^2 \mbI_M$. Then, we can express the log-likelihood $\ln \mathbb{P}(\bmby\vert\mbs)$, except the constant term in the case with prior channel knowledge, as $-\frac{1}{2} \lVert \bmby \rVert^2_{\mbGam_\cS^{-1}}$. Furthermore, $\mbGam_\cS^{-1}$ can be derived as follows:
\begin{align}
    \mbGam_\cS^{-1} &= \left(\sigma_w^2 \mbI_M + \mbPsi_\cS \mbK_\cS \mbPsi_\cS^H \right)^{-1}\\
    &= \frac{1}{\sigma_w^2}\left( \mbI_M - \mbPsi_\cS (\sigma_w^2 \mbK_\cS^{-1} + \mbPsi_\cS^H \mbPsi_\cS)^{-1} \mbPsi_\cS^H \right) \label{eq:inverseGam1}\\
    &\simeq \frac{1}{\sigma_w^2}\left( \mbI_M - \mbPsi_\cS (\mbPsi_\cS^H \mbPsi_\cS)^{-1} \mbPsi_\cS^H \right), \label{eq:inverseGam2}
\end{align}
where \eqref{eq:inverseGam1} is obtained from the matrix inversion lemma; \eqref{eq:inverseGam2} is derived under the assumption that $\sigma_w^2/[\mbK(\cS)]_{n,n} \ll \mbpsi_n^H \mbpsi_n$ for $n \in \cS$, providing mathematical tractability, as discussed in \cite{Masood2013}. Thus, $-\frac{1}{2} \lVert \bmby \rVert^2_{\mbGam_\cS^{-1}}$ can be approximated as $-\frac{1}{2 \sigma_w^2} \bmby \mbPi^\perp_\cS \bmby$ where $\mbPi^\perp_\cS = \mathbf{I} - \mbPsi_\cS(\mbPsi_\cS^H \mbPsi_\cS)^{-1} \mbPsi_\cS^H$. Therefore, by omitting the constant term, the likelihood $\mathbb{P}(\bmby \vert \cS)$ can be proportionally expressed as $\mathbb{P}(\bmby \vert \cS) \propto \exp \left( -\frac{1}{2 \sigma_w^2} \lVert \mbPi^\perp_\cS \bmby \rVert^2 \right)$. Consequently, the sparsity decision metric without prior channel knowledge can be expressed as
\begin{align} \label{eq:SparsityDecisionWithout}
    \alpha(\cS, \bmby) &= \ln \exp \left( -\frac{1}{2 \sigma_w^2} \lVert \mbPi^\perp_\cS \bmby \rVert^2 \right)\\
    &= \frac{1}{2 \sigma_w^2} \left\lVert \mbPsi_\cS(\mbPsi_\cS^H \mbPsi_\cS)^{-1} \mbPsi_\cS^H \bmby \right\rVert^2 - \frac{1}{2 \sigma_w^2} \lVert \bmby \rVert^2. \label{eq:SparsityDecisionWithout2}
\end{align}

In \eqref{eq:ChannelEstimateWithout} and \eqref{eq:SparsityDecisionWithout2}, the computation of $(\mbPsi_\cS^H \mbPsi_\cS)^{-1} \mbPsi_\cS^H \bmby$ is computationally intensive. However, we can express $(\mbPsi_\cS^H \mbPsi_\cS)^{-1} \mbPsi_\cS^H \bmby$ in a recursive form, making it computationally efficient. By calculating the inverse of $\mbPsi_\cS^H \mbPsi_\cS$ and simplifying using the block inversion formula \cite{Bernstein2009}, we can obtain the following expression:
\begin{align}
    (\mbPsi_\cS^H \mbPsi_\cS)^{-1} \mbPsi_\cS^H \bmby &=\! \left[ \begin{array}{cc}
    \mbPsi_{\ucS}^H \mbPsi_{\ucS} & \mbPsi_{\ucS}^H \mbPsi_i\\
    \mbPsi_i^H \mbPsi_{\ucS} & \mbPsi_i^H \mbPsi_i
    \end{array} \right]^{-1} \!\left[ \begin{array}{c}
    \mbPsi_{\ucS}^H \bmby\\
    \mbPsi_i^H \bmby
    \end{array} \right]\\
    &= \!\left[\! \begin{array}{c}
    \mbPsi_{\ucS}^\dagger \left( \mathbf{I} - \frac{1}{\mbPsi_i^H \mbPi^\perp_{\ucS} \mbPsi_i} \mbPsi_i \mbPsi_i^H \mbPi^\perp_{\ucS} \right) \bmby\\
    \frac{1}{\mbPsi_i^H \mbPi^\perp_{\ucS} \mbPsi_i} \mbPsi_i^H \mbPi^\perp_{\ucS} \bmby
    \end{array} \!\right]\!. \label{eq:BlockInverse}
\end{align}
Here, when $\cS = \ucS \cup \{i\}$, $\ucS$ represents the subset of $\cS$ obtained at the previous iteration. Furthermore, $i$ and $\mbPsi_i$ denote the index selected at the current iteration and the $i$-th column of $\mbPsi$, respectively. Additionally, we define $\mbPsi_{\ucS}^\dagger = (\mbPsi_{\ucS}^H \mbPsi_{\ucS})^{-1} \mbPsi_{\ucS}^H$ and $\mbPi^\perp_{\ucS} = \mathbf{I} - \mbPsi_{\ucS}\mbPsi_{\ucS}^\dagger$. Consequently, we can express $\alpha(\cS, \bmby)$ and $\mathbb{E}\left[\mbhAP \vert \cS, \bmby \right]$ recursively, enabling us to obtain them with lower computational complexity. The detailed process of the proposed algorithm without prior channel knowledge is summarized in Algorithm \ref{al:ChannelEstimation2}.

\begin{algorithm}[t]
\caption{BMP-based Hybrid-Field Scattering Channel Estimation without Prior Channel Knowledge} \label{al:ChannelEstimation2}
\DontPrintSemicolon
\SetKwProg{Init}{Initialize:}{}{}
\Init{$\cS = [\;]$}{
}
\For{$l = 1:L$}{
    \For{$n \in \{1:N'\}\setminus \cS$}{
        \eIf{$l=1$}{
            $\alpha_n^{(l)} = \frac{1}{2 \sigma_w^2} \left\lVert \mbPsi_\cS(\mbPsi_\cS^H \mbPsi_\cS)^{-1} \mbPsi_\cS^H \bmby \right\rVert^2 - \frac{1}{2 \sigma_w^2} \lVert \bmby \rVert^2$ for $\cS = \{n\}$
            }{
            Compute $\alpha_n^{(l)} = \alpha(\cS \cup \{n\}, \bmby)$ in \eqref{eq:SparsityDecisionWithout2} using \eqref{eq:BlockInverse}
            }
    }
    $n^* \gets$ index with the largest $\alpha_n^{(l)}$ \;
    $\cS \gets \cS \cup \{n^*\}$\;
    Update $\mathbb{E}\left[\mbhAP \vert \cS, \bmby \right]$ using \eqref{eq:BlockInverse} \;
}
$\hmbhAP = \mathbb{E}\left[\mbhAP \vert \cS, \bmby \right]$ and $\hmbh = \mbF \hmbhAP$
\end{algorithm}

In algorithms \ref{al:ChannelEstimation1} and \ref{al:ChannelEstimation2}, determining the exact value of $L$ can be challenging because we lack precise information about the true sparsity of the channel. To ensure an adequate number of nonzero elements in the sparse channel, it is advisable to set $L$ slightly higher than the anticipated true sparsity. In practice, one approach is to set a predefined threshold or limit for the search process, effectively constraining the search. This strategy necessitates having some approximate prior knowledge about the channel sparsity. For high-frequency systems, real-world measurements indicate that the typical number of the channel components is around 2 to 6, depending on the communication environments \cite{Sloane2023}. Therefore, it is a reasonable practice to set a fixed number of iterations for termination with a focus on identifying the most probable channel support rather than achieving absolute precision.

%%%%%%%%%%%%%%%%%%%%%%%%%%%%%%%%%%%%%%%%%%%%%%%%%%%%%%%%%%%%%%%%%%%
% % % % % % % % % % % % % % % % % % % % % % % % % % % % % % % % % %
%%%%%%%%%%%%%%%%%%%%%%%%%%%%%%%%%%%%%%%%%%%%%%%%%%%%%%%%%%%%%%%%%%%
\section{Numerical Results} \label{sec:NumericalResults}
%%%%%%%%%%%%%%%%%%%%%%%%%%%%%%%%%%%%%%%%%%%%%%%%%%%%%%%%%%%%%%%%%%%
% % % % % % % % % % % % % % % % % % % % % % % % % % % % % % % % % %
%%%%%%%%%%%%%%%%%%%%%%%%%%%%%%%%%%%%%%%%%%%%%%%%%%%%%%%%%%%%%%%%%%%

In this section, we assess the effectiveness of two proposed hybrid-field channel estimation methods, which are accompanied by the proposed ADMM-based pilot signal design: i) LoS channel estimation + Algorithm \ref{al:ChannelEstimation1} (``BMP w/ CSI'') and ii) LoS channel estimation + Algorithm \ref{al:ChannelEstimation2} (``BMP w/o CSI''). We compare these methods with the existing hybrid-field channel estimation algorithms, including hybrid-field OMP (``HF OMP'') \cite{Wei2022} and hybrid-field SD-OMP (``HF SD-OMP'')\cite{Hu2023}. Both the existing OMP-based hybrid-field channel estimation algorithms are greedy methods that sequentially identify the far-field channel and near-field channel components using the well-known OMP method. The hybrid-field SD-OMP incorporates additional rough support detection before the OMP-based channel estimation. It is noteworthy that the existing methods do not consider the LoS channel component although it is usually necessary in high frequency-band communications. Hence, in the numerical validation of these existing methods, we integrate the LoS channel component into the far-field channel components. Furthermore, the four aforementioned hybrid-field channel estimation algorithms evaluated in this section all require a similar level of computational complexity, approximately $\mathcal{O}((L+1)MN')$. Therefore, the performance evaluation is based on the normalized mean square error (NMSE), defined as $\mathrm{NMSE} = \mathbb{E}\left( \frac{\lVert \hmbh - \mbh \rVert^2}{\lVert \mbh \rVert^2} \right)$. In addition, we include a Genie-aided-LS algorithm, where perfect channel knowledge is available, serving as the lower bound for assessing the performance of these algorithms.
 
We consider a scenario with a pilot length of $M = 40$, $N = 128$ BS antenna elements and a carrier frequency of 50 GHz, corresponding to a Rayleigh distance of 49 meters, for performance validation. Furthermore, we set the minimum allowable distance of 4 meters between the BS and the scatterer and the maximum allowable distance of 60 meters between the BS and the user, and we also employ $Q = 4$ discrete distance sections for the polar domain transform matrix. Moreover, the angle and distance of the scatterers are uniformly and randomly generated. To obtain precise results, we perform 1,000 independent trials for each channel estimation algorithm. In the numerical results, the curves with circle, triangle, and plus sign represent the performance of HF OMP, HF SD-OMP, and Genie-aided-LS, respectively, while the curves with square and cross represent the proposed BMP w/ CSI and BMP w/o CSI, respectively. Furthermore, we define SNR as $P_\mbx/\sigma_w^2$ and specify that both far-field and near-field small scale channel gains $g_{l_f}$ and $g_{l_n}$   follow $\mathcal{CN}(0,1)$, i.e., $\sigma_\sfA^2 = \sigma_\sfP^2 = 1$, under the free-space path loss model as expressed in Eq. \eqref{eq:FFChannel} and \eqref{eq:NFChannel}. Additionally, the energy of the LoS channel component is set equal to the sum of the energies of all scattered channel components.

\begin{figure}[t] 
	\centering
	\includegraphics[width=0.9\columnwidth]{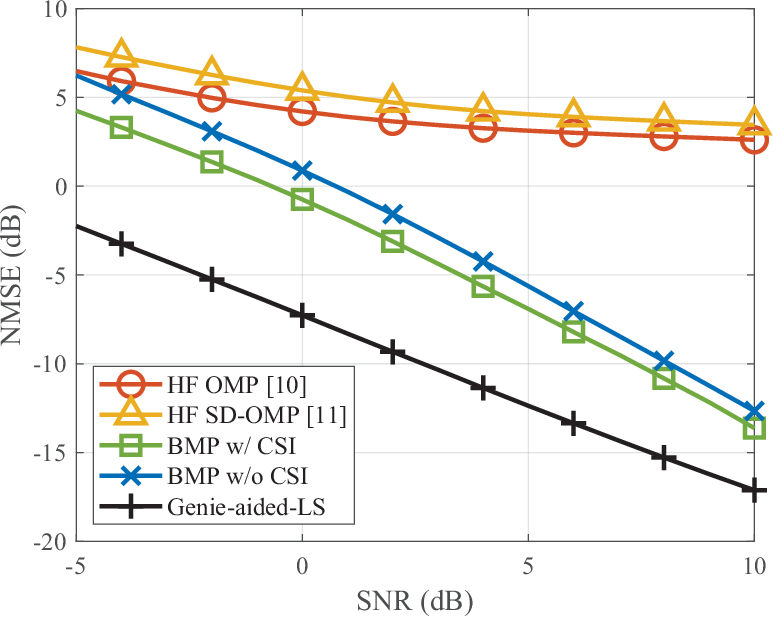}
	\caption{NMSE versus SNR with an LoS path and $L=4$ scattered paths for four hybrid-field channel estimation methods: 1) HF OMP; 2) HF SD-OMP; 3) BMP w/ CSI; 4) BMP w/o CSI.}
	\label{fig:SNR_NMSE}
\end{figure}

Fig. \ref{fig:SNR_NMSE} depicts a comparison of NMSE performance versus SNR with a total of 5 channel components. The figure shows that the performance of all algorithms improves as SNR increases. Remarkably, the proposed BMP w/ CSI consistently exhibits the best channel estimation performance across all SNR levels. Additionally, BMP w/o CSI demonstrates significantly improved NMSE performance compared to HF OMP and HF SD-OMP. Moreover, the two proposed algorithms approach Genie-aided-LS performance as SNR increases. However, the existing hybrid-field channel estimation algorithms, i.e., HF OMP and HF SD-OMP, consistently exhibit poor NMSE values above 0 dB for all SNR levels. It is worth noting that the two existing studies not only adopt a sequential method of estimating the far-field channel first and then the near-field channel but also do not consider pilot signal design at all for hybrid-field channel estimation. Due to the sequential estimation for both far-field and near-field channel components and the absence of pilot signal design, the two existing methods fail to deliver satisfactory hybrid-field channel estimation performance. Consequently, this result underscores the importance of co-designing pilot signal and channel estimator suitable for hybrid-field communications to achieve more reliable sparse channel recovery. Specifically, by satisfying the high SNR approximation in scenarios with high SNR, BMP w/o CSI can yield good recovery performance nearly as good as that of BMP w/ CSI. This implies that in situations with sufficient power for the pilot signal, fairly accurate hybrid-field channel estimation can be achieved without prior knowledge of the channel through BMP w/o CSI.

\begin{figure}[t] 
    \centering
    \includegraphics[width=0.9\columnwidth]{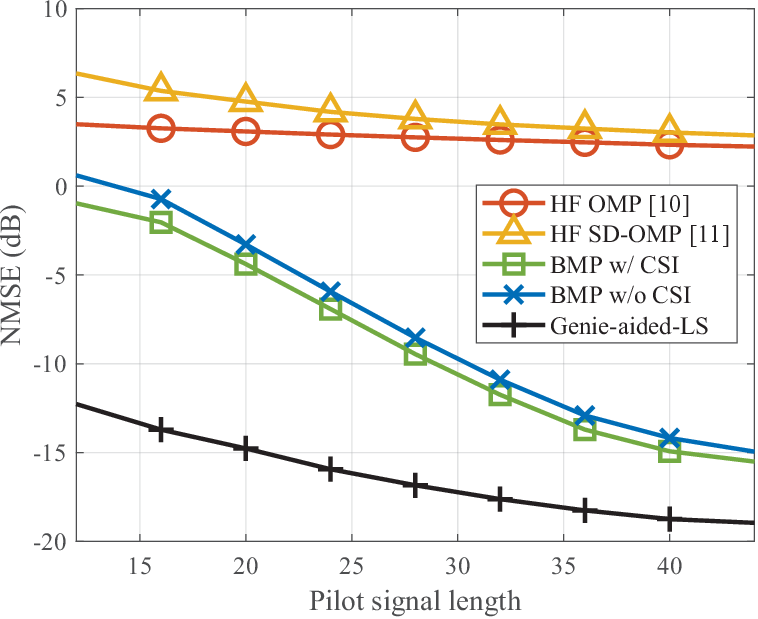}
    \caption{NMSE versus pilot signal length with an LoS path and $L=4$ scattered paths at an SNR of 10 dB for four hybrid-field channel estimation methods: 1) HF OMP; 2) HF SD-OMP; 3) BMP w/ CSI; 4) BMP w/o CSI.}
    \label{fig:Pilot_NMSE}
\end{figure}

Fig. \ref{fig:Pilot_NMSE} illustrates a comparison of NMSE performance versus pilot signal length $M$ with a total of 5 channel components at an SNR of 10 dB. As shown in Fig. \ref{fig:Pilot_NMSE}, the NMSE performance for four hybrid-field channel estimation algorithms and Genie-aided-LS generally improves as pilot signal length $M$, corresponding to the number of observations in the CS system, increases. With the increase in pilot signal length, our two proposed algorithms approach the performance of Genie-aided-LS. Furthermore, at high SNR, the two proposed algorithms demonstrate nearly the same NMSE performance, adhering to the high SNR approximation in BMP w/o CSI. Moreover, it is evident that the two proposed algorithms can yield reliable sparse channel recovery with a small number of pilot signals, approximately $M = 35$. In contrast, the conventional algorithms show minimal performance improvement even as the pilot signal length increases. Consequently, under more sparse channel environments resembling real hybrid-field channel communications, Figs. \ref{fig:SNR_NMSE} and \ref{fig:Pilot_NMSE} demonstrate that while the existing methods yield unreliable results, the proposed algorithms exhibit outstanding channel estimation performance close to the optimal method.

\begin{figure}[t]
    \centering
    \includegraphics[width=0.9\columnwidth]{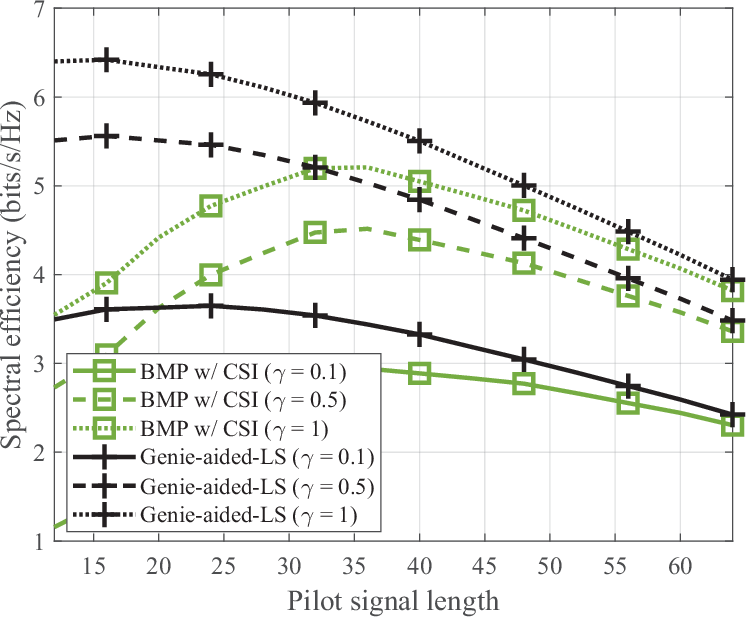}
    \caption{Spectral efficiency versus pilot signal length, considering variations in transmission power with an LoS path and $L=4$ scattered paths at an SNR of 10 dB for BMP w/ CSI.}
    \label{fig:Pilot_SE}
\end{figure}

Fig. \ref{fig:Pilot_SE} illustrates spectral efficiency versus pilot signal length, considering variations in data transmission power, at an SNR of 10 dB for the pilot sequence and a total of 5 channel components for BMP w/ CSI\footnote{Given the two proposed algorithms exhibit nearly the same NMSE performance at an SNR of 10 dB in Fig. \ref{fig:Pilot_NMSE}, we specifically address spectral efficiency for BMP w/ CSI.}. In the figure, spectral efficiency represents the achievable data rate, considering data transmission power, and pilot transmission power and duration required for channel estimation. Here, $\gamma$ stands for the ratio of data transmission power to pilot transmission power, with the general condition that data transmission power is smaller than or equal to pilot transmission power \cite{Park2022}. Furthermore, we consider a fast-moving user with 20 m/s at a 240 kHz subcarrier spacing, implying a channel coherence time of 0.5 ms and a symbol duration of 4.17 $\mu$s \cite{Lu2020}. As depicted in the figure, for all $\gamma$ values, the proposed BMP w/ CSI can achieve spectral efficiency close to the optimal method, i.e., the Genie-aided-LS, with only about $M = 40$ pilot signals. Additionally, the spectral efficiencies of the proposed algorithm for each $\gamma$ with $M = 40$ do not significantly differ from the best performance of the Genie-aided-LS algorithm for the corresponding $\gamma$. It is worth noting that since more pilot signals can be available with lower pilot overhead under more static environments or wider subcarrier spacing, implying longer channel coherence time or shorter symbol duration, it is anticipated that the hybrid-field channel estimation performance can be further enhanced. Consequently, these results demonstrate that our approaches are sufficiently feasible, and the proposed algorithm can yield fairly good sparse channel recovery, considering practical communication scenarios in hybrid-field XL-MIMO systems.

\begin{figure}[t] 
    \centering
    \includegraphics[width=0.9\columnwidth]{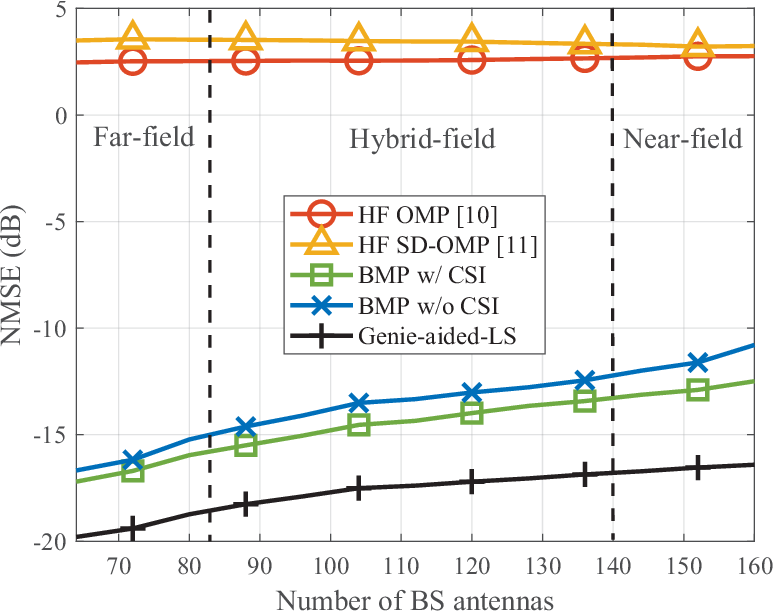}
    \caption{NMSE versus the number of BS antennas, with an LoS path and $L=4$ scattered paths at an SNR of 10 dB, for four hybrid-field channel estimation methods: 1) HF OMP; 2) HF SD-OMP; 3) BMP w/ CSI; 4) BMP w/o CSI.}
    \label{fig:Antenna_NMSE}
\end{figure}

Fig. \ref{fig:Antenna_NMSE} illustrates a comparison of NMSE performance concerning the number of BS antennas, considering an LoS channel path and $L=4$ scattered channel paths, at an SNR of 10 dB. The scenario involves 4 fixed scatterers uniformly and randomly distributed within the communication coverage of the BS, and over 1000 independent trials, we assess NMSE performance as the number of BS antennas varies. The figure shows that with fewer BS antennas, the Rayleigh distance is considerably small, placing all scatterers in the far-field region. However, as the number of antennas increases, the Rayleigh distance also increases, transitioning the far-field region into a near-field region. Consequently, since the scatterers are located fixed in the considered scenario, the communication area changes from the far-field region to the hybrid-field region and then to the near-field region as the number of BS antennas increases. As shown in the figure, sparse channel recovery performance tends to degrade due to the nature of CS as the number of BS antennas increases \cite{Foucart2013}. Furthermore, the figure demonstrates that the two proposed algorithms significantly improve channel estimation accuracy compared to the existing hybrid-field channel estimation methods and are fairly comparable to the Genie-aided-LS method. Consequently, these results affirm that the proposed algorithms are well-suited for different numbers of BS antennas in a predefined communication scenario.

\begin{figure}[t] 
    \centering
    \includegraphics[width=0.9\columnwidth]{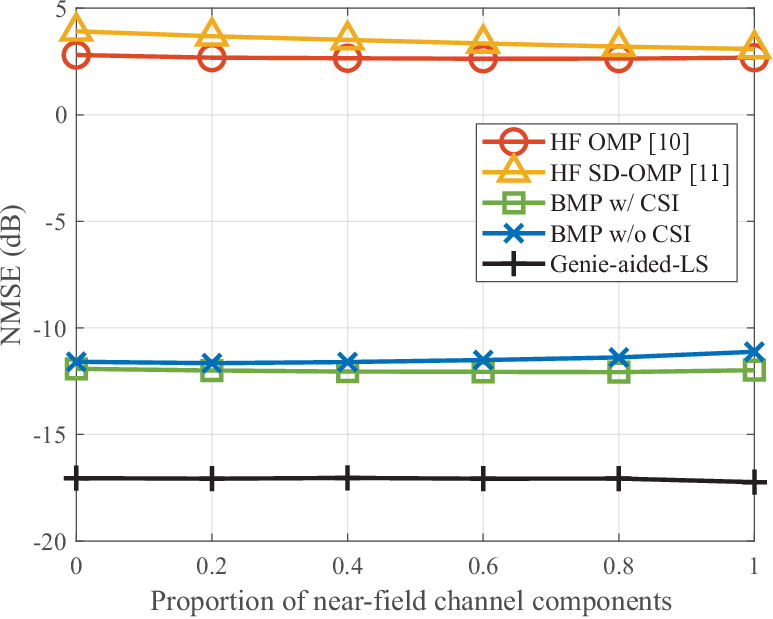}
    \caption{NMSE versus proportion of near-field channel components, with an LoS path and $L=5$ scattered paths at an SNR of 10 dB, for four hybrid-field channel estimation methods: 1) HF OMP; 2) HF SD-OMP; 3) BMP w/ CSI; 4) BMP w/o CSI.}
    \label{fig:Ratio_NMSE}
\end{figure}

Fig. \ref{fig:Ratio_NMSE} illustrates a comparison of NMSE performance concerning the proportion of near-field channel components relative to the total scattered channel components. As shown in the figure, our two proposed algorithms consistently outperform the existing hybrid-field channel estimation algorithms across all ratios of near-field channel components. As mentioned earlier, in contrast to the existing algorithms, the proposed algorithms incorporate additional pilot design for hybrid-field communications and joint hybrid-field scattering channel estimation, resulting in superior NMSE performance that surpasses the existing algorithms and approaches that of the optimal method. Specifically, despite the existing algorithms having prior information about the ratio of near-field channel components, our proposed algorithms, lacking such prior knowledge, demonstrate superior performance. Therefore, these results indicate that our two proposed algorithms perform well for all ratios between the two scattered channel components, even in the absence of prior knowledge about the ratio. Consequently, Figs. \ref{fig:Antenna_NMSE} and \ref{fig:Ratio_NMSE} demonstrate that our proposed algorithms are well-designed for all scenarios in XL-MIMO systems, including scenarios with only far-field or near-field channel components, as well as the hybrid-field scenario.

%%%%%%%%%%%%%%%%%%%%%%%%%%%%%%%%%%%%%%%%%%%%%%%%%%%%%%%%%%%%%%%%%%%
% % % % % % % % % % % % % % % % % % % % % % % % % % % % % % % % % %
%%%%%%%%%%%%%%%%%%%%%%%%%%%%%%%%%%%%%%%%%%%%%%%%%%%%%%%%%%%%%%%%%%%
\section{Conclusion} \label{sec:Conclusion}
%%%%%%%%%%%%%%%%%%%%%%%%%%%%%%%%%%%%%%%%%%%%%%%%%%%%%%%%%%%%%%%%%%%
% % % % % % % % % % % % % % % % % % % % % % % % % % % % % % % % % %
%%%%%%%%%%%%%%%%%%%%%%%%%%%%%%%%%%%%%%%%%%%%%%%%%%%%%%%%%%%%%%%%%%%

In this paper, we addressed hybrid-field channel estimation in XL-MIMO systems. We formulated the hybrid-field channel estimation problem and tackled two key challenges for hybrid-field channel estimation, To resolve these challenges, we proposed the ADMM-based pilot signal design algorithm to minimize the mutual coherence of the sensing matrix in the CS problem. Then, building upon the proposed pilot signals, we further developed the two-stage hybrid-field channel estimation algorithm, incorporating LoS channel estimation and hybrid-field scattering channel estimation. Through experimental results, we validated that our co-design of the pilot signal and channel estimator for the hybrid-field channel outperforms the existing methods in terms of NMSE performance. This result underscores the effectiveness of our proposed approach in enhancing the accuracy and reliability of hybrid-field channel estimation in XL-MIMO systems.

\ifCLASSOPTIONcaptionsoff
\newpage
\fi

\bibliographystyle{IEEEtran}
\bibliography{Hybrid_CE}

\end{document}